\documentclass[journal]{IEEEtran}

\IEEEoverridecommandlockouts

\usepackage[cmex10]{amsmath}
\usepackage{amssymb}

\usepackage{cases}
\usepackage{multirow}
\usepackage{empheq}

\usepackage{units}
\usepackage[noadjust]{cite}
\usepackage{verbatim}

\usepackage{algorithm}
\usepackage{algpseudocode}

\usepackage{siunitx}

\ifCLASSINFOpdf
\usepackage[pdftex]{graphicx}
\else
\usepackage[dvips]{graphicx}
\fi
\usepackage[caption=false,font=footnotesize]{subfig}

\usepackage{color}
\usepackage{multirow}

\usepackage[hyphens]{url}

\newtheorem{theorem}{Theorem}
\newtheorem{lemma}[theorem]{Lemma}

\newtheorem{corollary}[theorem]{Corollary}
\newtheorem{definition}[theorem]{Definition}

\newtheorem{proposition}[theorem]{Proposition}
\newtheorem{remark}[theorem]{Remark}

\newcommand{\vect}[1]{\mathbf{#1}}

\begin{document}
	
	\sloppy
	
	\title{Generalized LRS Estimator for \\ Min-entropy Estimation}
	
	\author{
		\IEEEauthorblockN{Jiheon Woo, Chanhee Yoo, Young-Sik Kim, Yuval Cassuto, and Yongjune Kim}	\\		
		\thanks{J. Woo and C. Yoo contributed equally. J. Woo, C. Yoo, and Y. Kim are with the Department of Information and Communication Engineering, Daegu Gyeongbuk Institute of Science and Technology (DGIST), Daegu 42988, South Korea (e-mail: \{jhwoo1997, yoo1209, yjk\}@dgist.ac.kr). Y.-S. Kim is with the Department of Information and Communication Engineering, Chosun University, Gwangju 61452, South Korea (e-mail: iamyskim@chosun.ac.kr). Y. Cassuto is with the Viterbi Department of Electrical and Computer Engineering, Technion--Israel Institute of Technology, Haifa 32000, Israel (e-mail: ycassuto@ee.technion.ac.il).}		
	}
	
	
	\maketitle
	
	\begin{abstract}
	
    The min-entropy is a widely used metric to quantify the randomness of generated random numbers, which measures the difficulty of guessing the most likely output. It is difficult to accurately estimate the min-entropy of a non-independent and identically distributed (non-IID) source. Hence, NIST Special Publication (SP) 800-90B adopts ten different min-entropy estimators and then conservatively selects the minimum value among ten min-entropy estimates. Among these estimators, the longest repeated substring (LRS) estimator estimates the collision entropy instead of the min-entropy by counting the number of repeated substrings. Since the collision entropy is an upper bound on the min-entropy, the LRS estimator inherently provides \emph{overestimated} outputs. In this paper, we propose two techniques to estimate the min-entropy of a non-IID source accurately. The first technique resolves the overestimation problem by translating the collision entropy into the min-entropy. Next, we generalize the LRS estimator by adopting the general R{\'{e}}nyi entropy instead of the collision entropy (i.e., R{\'{e}}nyi entropy of order two). We show that adopting a higher order can reduce the variance of min-entropy estimates. By integrating these techniques, we propose a generalized LRS estimator that effectively resolves the overestimation problem and provides stable min-entropy estimates. Theoretical analysis and empirical results support that the proposed generalized LRS estimator improves the estimation accuracy significantly, which makes it an appealing alternative to the LRS estimator. 
   
	\end{abstract}  
	
	\section{Introduction}
	
	Random numbers are essential for generating cryptographic information such as secret keys, nonces, salt values, \emph{etc}. The security of cryptographic systems crucially relies on the randomness of the generated random numbers~\cite{Turan2018recommendation,Hagerty2012entropy,Kelsey2015predictive,Amaki2013worst}. Hence, it is critical to quantify the randomness of the generated numbers accurately. Among several ways to quantify randomness of generated random numbers, \emph{entropies} are widely used metrics in standards such as AIS.31~\cite{Killmann2011proposal}, NIST Special Publication (SP) 800-22~\cite{Rukhin2010statistical}, and NIST SP 800-90B~\cite{Turan2018recommendation}.
	
	There are several kinds of entropies such as Shannon entropy, R{\'{e}}nyi entropy, and min-entropy. Among them, the min-entropy is a well-justified metric in cryptographic applications~\cite{Turan2018recommendation,Kelsey2015predictive} since the min-entropy measures the difficulty of guessing the most likely output. Furthermore, the min-entropy is a lower bound on the Shannon entropy and the R{\'{e}}nyi entropy, i.e., one of the most conservative metrics. 
	
	For independent and identically distributed (IID) sources, the min-entropy can be readily estimated by the empirical estimator~\cite{Turan2018recommendation}. However, it is difficult to estimate the min-entropy of non-IID sources accurately. Hence, NIST SP 800-90B proposes ten different min-entropy estimators for non-IID sources (see Table~\ref{tab:estimators}). These estimators independently perform their own estimations based on different statistics of the examined non-IID sources. Then, NIST SP 800-90B conservatively selects the minimum among these ten different values as the final estimate of min-entropy.  

	
	Among the ten min-entropy estimators, the \emph{longest repeated substring (LRS) estimator} estimates the collision entropy (the R{\'{e}}nyi entropy of order two) based on the number of repeated substrings, i.e., collision counts~\cite{Turan2018recommendation}. Since the collision entropy is an upper bound on the min-entropy, the LRS estimator overestimates the min-entropy, which violates the conservative estimation of NIST SP 800-90B. NIST SP 800-90B selects the minimum among ten estimates as the final min-entropy estimate; hence, the overestimated value by the LRS estimator would typically not affect the final estimate, which could undermine the justification to include the LRS estimator in NIST SP 800-90B. 
	
	In this paper, we propose two techniques to amend the LRS estimator for accurate min-entropy estimation. The first technique resolves the overestimation problem by enabling the estimation of the min-entropy instead of the collision entropy. The proposed technique leverages the inequality of \cite[Theorem 6]{Ben-Bassat1978renyi}, which characterizes the relation between the min-entropy and the R{\'{e}}nyi entropy. For this technique, we show that the proposed estimator is almost unbiased for binary sources, which are the most common sources. Next, we generalize the LRS estimator by parameterizing the order, i.e., $\alpha$ of the R{\'{e}}nyi entropy. By adopting a higher order $\alpha$ than two of the collision entropy, the variance of min-entropy estimates can be reduced, which leads to more stable estimates. We analytically show that the variance of estimates decreases with $\alpha$, although the reduction of the variance diminishes as $\alpha$ increases. 
	
	By integrating these two techniques, we propose a \emph{generalized} LRS estimator that improves the estimation accuracy by twofold: 1) the bias is reduced by resolving the overestimation problem of the LRS estimator; 2) the variance of the min-entropy estimates is reduced by adopting the higher order of the R{\'{e}}nyi entropy. Theoretical analysis and empirical results support that the generalized LRS estimator significantly improves the estimation accuracy of the LRS estimator, although both the LRS estimator and the proposed estimator rely on the same statistics, i.e., counts of repeated substrings. We believe that the generalized LRS estimator is an appealing alternative to the LRS estimator of NIST SP 800-90B since the generalized LRS estimator outputs a more accurate and stable min-entropy estimate from the same statistics of a given sequence. 
	
	\begin{table}[!t]
	\renewcommand{\arraystretch}{1.2}
	\caption{Classification of NIST SP 800-90B Estimators~\cite{Turan2018recommendation,Zhu2020analysis}}
	\vspace{-2mm}
	\label{tab:estimators}
	\centering
	\begin{tabular}{|c|c|}	\hline
		Statistic-based estimator~\cite{Hagerty2012entropy} & Prediction-based estimator~\cite{Kelsey2015predictive} \\ \hline \hline
		Most common value estimator  &  MultiMCW prediction estimator \\ 
		Collision estimator & Lag prediction estimator \\ 
		Markov estimator & MultiMMC prediction estimator \\ 
		Compression estimator & LZ78Y prediction estimator \\ 
		$t$-Tuple estimator &  \\ 
		LRS estimator  & \\ \hline			
	\end{tabular}
	\vspace{-2mm}
	\end{table} 		
	
	The rest of this paper is organized as follows. Section~\ref{sec:prelim} briefly explains the several types of entropies and the LRS estimator of NIST SP 800-90B. Section~\ref{sec:proposed_improved} presents the improved LRS estimator that accurately estimates the min-entropy. Section~\ref{sec:proposed_generalized} proposes the generalized LRS estimator that enables more stable estimation. Section~\ref{sec:numerical} provides numerical results and Section~\ref{sec:conclusion} concludes. 
		
	\section{Preliminaries: Entropies and LRS Estimator}\label{sec:prelim}	
	
	\subsection{Entropies and Power Sum} \label{sec:entropies}
	
	Suppose that the input sequence $\vect{s} = (s_1,\ldots,s_L)$, where $s_i \in \{x_1, \ldots, x_k\}$ is generated from a given source $S$. The Shannon entropy is defined as
	\begin{equation} \label{eq:Shannon_entropy}
	    H(S) = H(\vect{p}) = - \sum_{i = 1}^{k}{p_i \log_2{p_i}},
	\end{equation}
	where $\vect{p} = (p_1, \ldots, p_k)$ denotes the distribution of $S$.
	
	The R{\'{e}}nyi entropy of order $\alpha \in (0,1) \cup (1,\infty)$ is defined as
	\begin{equation} \label{eq:Binary_Renyi_entropy}
	H_{\alpha}(S) = H_{\alpha}(\vect{p}) =\frac{1}{1 - \alpha} \log_2\sum_{i=1}^k p_i^\alpha. 
	\end{equation} 
    For $\alpha = 2$, the R{\'{e}}nyi entropy corresponds to the \emph{collision} entropy $H_{2}(S)$ as follows: 
    \begin{equation} \label{eq:collision_entropy}
	H_{2}(S) = H_{2}(\vect{p}) =-\log_2{\sum_{i=1}^{k}{p_i^2}}.
	\end{equation}
     
    The min-entropy is defined as
	\begin{align} \label{eq:min_entropy}
	H_{\infty}(S) &= H_{\infty}(\vect{p}) = -\log_2{\theta}, 
	\end{align} 
	where 
	\begin{equation} \label{eq:theta}
	    \theta = \max_{i \in \{1,\ldots,k\}} \{p_i\}. 
	\end{equation}
	
	\begin{definition}[Power Sum] \label{def:power_sum}  
	The power sum of order $\alpha$ (i.e., the $\alpha$th moment) for a distribution $\vect{p}$ is defined as
    \begin{equation} \label{eq:power_sum}
		M_{\alpha} (\vect{p}) = \sum_{i = 1}^{k}{p_i^\alpha}.
	\end{equation}
	\end{definition}
	
	\begin{remark}[Collision Probability] \label{rem:power_sum_collision}
	    The power sum of order $\alpha = 2$ (i.e., $M_{2}(\vect{p})$) is equivalent to the collision probability, which is the probability that two arbitrary source outputs are equal. 
	\end{remark}

	\begin{remark} \label{rem:power_sum_entropy}
	    The R{\'{e}}nyi entropy of order $\alpha$ is $H_{\alpha}(\vect{p}) = \frac{1}{1-\alpha}\log_2{M_{\alpha}(\vect{p})}$. 
	\end{remark}

	\begin{remark} \label{rem:entropy_relation}
		The following relations are well known:
		\begin{align}
		H(S) &= \lim_{\alpha \rightarrow 1} H_{\alpha}(S),  \\
		H_{\infty}(S) &= \lim_{\alpha \rightarrow \infty} H_{\alpha}(S). \label{eq:min_Renyi}
		\end{align}
	\end{remark}
   
	\begin{remark} \label{rem:Renyi}
		The R{\'{e}}nyi entropy is non-increasing in $\alpha$~\cite{Beck1993thermodynamics}. Hence, $\forall \alpha, H_{\infty}(S) \le H_{\alpha}(S)$, i.e., the min-entropy is a lower bound on the Shannon entropy and the R{\'{e}}nyi entropy.
	\end{remark}
	
	\subsection{LRS Estimator and Its Overestimation Problem}
	
	\begin{algorithm}[!t] 
		\caption{LRS estimator of NIST 800-90B~\cite{Turan2018recommendation}} \label{algo:nist}
		\textbf{Input:} Sequence $\vect{s} = (s_1,\ldots,s_L)$ where $s_i \in \{x_1, \ldots,x_k\}$. \\
		\textbf{Output:} Collision entropy $H_{2}(S)$.
		\begin{algorithmic}[1]
			\State Find the smallest $u$ such that the number of occurrences of the most common $u$-tuple in $\vect{s}$ is less than 35.
			\State Find the largest $v$ such that the number of occurrences of the most common $v$-tuple in $\vect{s}$ is at least 2. \Comment{Longest repeated substring problem}
			\For {$w \in \{u, u+1, \ldots,v\}$}
			\State Estimate the estimated $w$-tuple collision probability: 
			\begin{equation} \label{eq:nist_collision}
			    P_w := \frac{\sum_{i} \binom{C_i}{2} }{\binom{l}{2}},
			\end{equation}
		    	where $C_i$ is the number of occurrences of the $i$th unique $w$-tuple and $l$ is the total number of $w$-tuples.
			\State Compute the collision probability per sample: 
			\begin{equation} \label{eq:nist_collision_norm}
			    \widetilde{P}_{w}:=P_w^{1/w}.
			\end{equation} 
			\EndFor
			\State $\widehat{p}_{c}:=\max \left\{\widetilde{P}_{u}, \ldots,\widetilde{P}_{v} \right\}$.
			\State $\widetilde{p}_c := \min \left\{1,\widehat{p}_c+2.576\sqrt{\frac{\widehat{p}_c(1 - \widehat{p}_c)}{L-1}} \right\}$.  
			\State $H_{2}(S):=-\log_2{\widetilde{p}_c}$.
		\end{algorithmic}
	\end{algorithm}	
   
    For non-IID sources, NIST SP 800-90B proposes ten different min-entropy estimators (see Table~\ref{tab:estimators}). These estimators independently perform their own estimations based on different statistics calculated from the examined non-IID sources. Among these ten estimators, the $t$-tuple estimator and the LRS estimator compute entropies based on the frequency of substrings (tuples) in the input sequence $\vect{s}$. The $t$-tuple estimator estimates the min-entropy based on the frequency of some fixed-length repeated substrings. The LRS estimator handles substring sizes that are too large for the $t$-tuple estimator~\cite{Turan2018recommendation,Zhu2020analysis}. 
   
    Algorithm~\ref{algo:nist} describes the LRS estimator in NIST SP 800-90B. Step 1 finds the smallest $u$ such that the number of occurrences of the most common $u$-tuple is less than 35. Step 2 solves the well-known \emph{longest repeated substring problem} and set $v$ as its length. Then, the range of $w$ becomes $\{u, u+1, \ldots, v\}$. In contrast, the $t$-tuple estimator finds the largest $t$ such that the number of occurrences of the most common $t$-tuple is at least 35 and the range of $w$ in that test equals $\{1,\ldots, t\}$ where $t < u$. Note that the $t$-tuple estimator and the LRS estimator calculate the entropies based on disjoint substring lengths, where the LRS estimator handles the longer substrings. Hence, the $t$-tuple estimator and the LRS estimator are \emph{complementary}. 
   
    The LRS estimator estimates collision entropy instead of the min-entropy. Step 4 calculates the empirical collision probability of length-$w$ substrings. The LRS estimator of NIST SP 800-90B uses \emph{overlapped} tuples, i.e., $l = L - w + 1$ in Step 4. For \emph{non-overlapped} tuple counts, the total number of $w$-tuples becomes $l = \left\lfloor \frac{L}{w} \right\rfloor$. 
    
    The collision probability estimation by \eqref{eq:nist_collision} is a key step of the LRS estimator, which was considered in~\cite{Goldreich2000testing,Batu2013testing} for testing whether a distribution is close to uniform. Note that \eqref{eq:nist_collision} is an \emph{unbiased} estimator of the collision probability~\cite{Acharya2015complexity}. 
    
    Step 5 computes the collision probability per sample (to normalize the estimated entropy) and Step 7 conservatively chooses the maximum (across $w$) collision probability (i.e., the minimum collision entropy). Step 8 ensures the confidence level of \unit[99]{\%} under the Gaussian assumption. Although Step 7 and Step 8 follow the conservative approach of NIST SP 800-90B, the LRS estimator \emph{overestimates} the min-entropy since Step 9 estimates the collision entropy instead of the min-entropy.
    
\begin{figure}[t]
		\centering
		\includegraphics[width=0.45\textwidth]{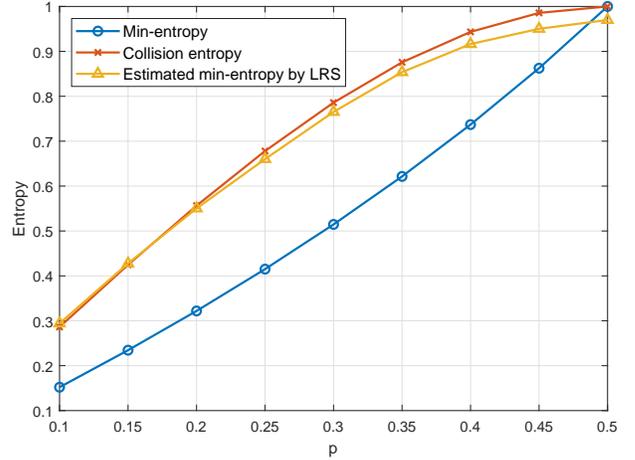}
		\caption{Comparison of min-entropy, collision entropy and estimated value by LRS estimator for binary memoryless source (BMS) with parameter $p$. While the LRS estimator has a small underestimation gap to the true collision entropy (due to Step 8 in Algorithm~\ref{algo:nist}), its overestimation gap (bias) to the min-entropy is significant.}
		\label{fig:lrs_estimated}
	\end{figure}

    Fig.~\ref{fig:lrs_estimated} shows the ramifications from the fact that the LRS estimator estimates the collision entropy instead of the min-entropy. The bias between the actual min-entropy and the estimate by the LRS estimator is considerable except for $p = 0.5$. The numerical results in~\cite[Table 2 and 3]{Zhu2017analysis} also confirm this overestimation problem for the first-order Markov source and several pseudo-random data. 
    
    NIST SP 800-90B conservatively selects the minimum among estimated values by ten estimators. Hence, an overestimated value by the LRS estimator would not affect the final estimate. Since the LRS estimator computes based on larger sizes of susbtrings than the $t$-tuple estimator, the LRS estimator takes around ten times longer execution time than the $t$-tuple estimator~\cite[Table V]{Zhu2020analysis}. In spite of this considerable execution time, the LRS estimator rarely affects the final min-entropy estimate due to overestimation problem noted above.

	
	\section{Min-entropy Estimation by LRS Estimator}\label{sec:proposed_improved}
		
	\subsection{Min-entropy Estimation by LRS Estimator} \label{sec:proposed_improved_scheme}
	
	
	In this section, we propose a method to resolve the overestimation problem of the LRS estimator. The proposed method attempts to estimate the min-entropy instead of the collision entropy by using the estimation values of the LRS estimator and the following bound. 
	
	\begin{lemma}[{\cite[Theorem 6]{Ben-Bassat1978renyi}}]\label{lem:Fano_Renyi} 
	Suppose that $\theta = \max_{i\in\{1,\ldots,k\}}\{p_i\}$. Then, the following inequality holds:   
		\begin{equation} \label{eq:Fano_Renyi}
		H_{\alpha}(S) \le \frac{1}{1 - \alpha} \log_2{\left(\theta^{\alpha} +  \frac{(1-\theta)^\alpha}{(k-1)^{\alpha-1}}\right)} 
		\end{equation} 
		for $\alpha \ne 1$. The bound is achieved with equality by the \emph{near-uniform} distribution $\vect{p}_{\text{NU}}(\theta) = (p_1, \ldots, p_k)$ where
		\begin{equation} \label{eq:near_uniform}
		    p_i = 
		    \begin{cases}
			\ \theta, & \text{if}~i=1; \\
			\frac{1-\theta}{k-1}, & \text{otherwise}. \\
		    \end{cases}
        \end{equation}
        Without loss of generality, $p_1 \ge \ldots \ge p_k$ is assumed. 
	\end{lemma}
	
	The bound \eqref{eq:Fano_Renyi} is the counterpart of Fano's inequality, which applies to the Shannon entropy.
	
	\begin{theorem} \label{thm:theta_ub} For the estimated collision probability $\widehat{p}_c$ by Algorithm~\ref{algo:nist}, the following inequality holds: 
	\begin{equation} \label{eq:theta_ub}
	    \theta \le \frac{\sqrt{(k-1)({p}_c k - 1)}+1}{k},
	\end{equation}
	where $p_c = \mathbb{E}(\widehat{p}_c)$. Since the near-uniform distribution achieves \eqref{eq:Fano_Renyi} with equality, \eqref{eq:theta_ub} is the \emph{sharp} upper bound. 
	\end{theorem}
	\begin{IEEEproof} 
	    For $\alpha > 1$, \eqref{eq:Fano_Renyi} leads to 
	    \begin{equation} \label{eq:collision_bound}
	        M_{\alpha}(\vect{p}) \ge \theta^\alpha +  \frac{(1-\theta)^\alpha}{(k-1)^{\alpha-1}}. 
	    \end{equation}
	    Since $M_{2}(\vect{p})$ is equivalent to the collision probability $p_c$~\cite{Acharya2015complexity}, we can set $p_c \ge \theta^2 +  \frac{(1-\theta)^2}{(k-1)}$. By \eqref{eq:theta}, it is clear that $\theta \ge \frac{1}{k}$. Since $\theta^2 + \frac{(1-\theta)^2}{(k-1)}$ is a non-decreasing function of $\theta$ for $\theta \ge \frac{1}{k}$, \eqref{eq:theta_ub} holds. 
	\end{IEEEproof}
	
	Based on Theorem~\ref{thm:theta_ub}, we estimate $\widehat{\theta}$ as follows:
	\begin{equation} \label{eq:theta_est}
	    \widehat{\theta} = \frac{\sqrt{(k-1)(\widehat{p}_c k-1)}+1}{k},
	\end{equation}
	which is a conservative min-entropy estimation. It is because an upper bound on $\theta$ leads to a lower bound on $H_{\infty}(S)$, i.e., 
	\begin{equation} \label{eq:min_entropy_ub}
	    H_{\infty}(S) = - \log_2{\theta} \ge -\log_2{\widehat{\theta}} 
    \end{equation}
    if $\widehat{p}_c = p_c$. 
	
	\begin{algorithm}[!t] 
		\caption{Proposed Estimator (Improved LRS Estimator)} \label{algo:improved_lrs}
		\textbf{Input:} Sequence $\vect{s}=(s_1,\ldots,s_L)$ where $s_i \in \{x_1,\ldots,x_k\}$. \\
		\textbf{Output:} Min-entropy $H_{\infty}(S)$.
		\begin{algorithmic}[1] 
			\State Estimate $\widehat{p}_c$ from $\vect{s}$ by Algorithm~$\ref{algo:nist}$.
			\If{$\widehat{p}_c > \frac{1}{k}$}
			    \State $\widehat{\theta} := \frac{\sqrt{(k-1)(\widehat{p}_c k-1)}+1}{k}$.
	        \Else
	            \State $\widehat{\theta} := \frac{1}{k}$.
            \EndIf
		    \State $\widetilde{\theta}:=\min\left(1,\widehat{\theta}+2.576\sqrt{\frac{\widehat{\theta}(1-\widehat{\theta})}{L-1}}\right)$.
			\State $H_{\infty}(S):=-\log_2{\widetilde{\theta}}$.
		\end{algorithmic}
	\end{algorithm}		

    Algorithm~\ref{algo:improved_lrs} describes the proposed min-entropy estimator. Algorithm~\ref{algo:improved_lrs} enables estimating the min-entropy instead of the collision entropy by leveraging the LRS estimator. Step 1 of Algorithm~\ref{algo:improved_lrs} estimates the collision probability by using Algorithm~\ref{algo:nist}. Theoretically, $p_c \ge \frac{1}{k}$ where the equality is achieved by the uniform distribution. If $\widehat{p}_c < \frac{1}{k}$, then we know that it results from estimation errors. Hence, in this case we set $\widehat{p}_c = \frac{1}{k}$, which leads to $\widehat{\theta} = \frac{1}{k}$. Step 7 ensures the confidence level of \unit[99]{\%} as in Step 8 of Algorithm~\ref{algo:nist}. 
    
    The proposed estimator attempts to estimate a \emph{lower} bound on the min-entropy whereas the LRS estimator estimates an \emph{upper} bound on the min-entropy (i.e., collision entropy). The proposed estimator matches the conservative approach of NIST SP 800-90B. It is worth mentioning that the collision estimator and the compression estimator of NIST SP 800-90B (Table~\ref{tab:estimators}) also estimate lower bounds on the min-entropy by using the near-uniform distribution as in the proposed estimator. Hence, the proposed estimator is better aligned with other estimators of NIST SP 800-90B than the LRS estimator. 
    
    Importantly, the proposed estimator is \emph{unbiased} for binary sources (i.e., it estimates the min-entropy itself instead of the lower bound since any binary distributions are near-uniform). In the next subsection, we further investigate the proposed estimator's bias properties. 
    
	\subsection{Bias of Proposed Estimator \label{sec:proposed_improved_bias}}
	
	We investigate the biases of the conventional LRS estimator and the proposed estimator. For the analysis, we neglect the step for \unit[99]{\%} confidence interval. Hence, $\widehat{p}_c$ and $\widehat{\theta}$ instead of $\widetilde{p}_c$ and $\widetilde{\theta}$ are considered in our analysis.   
	
	The bias of the LRS estimator is given by
	\begin{align} \label{eq:bias_lrs}
	    b_{\text{LRS}}(S) & = \mathbb{E}(- \log_2{\widehat{p}_c}) - H_{\infty}(S). 
	\end{align}
	
	\begin{proposition}
	    The LRS estimator is \emph{overestimating}, i.e., $b_{\text{LRS}}(S) > 0$. 
	\end{proposition}
	\begin{IEEEproof}
	We show that $b_{\text{LRS}}(S) > 0$ as follows:
	\begin{align}
	    b_{\text{LRS}}(S) & = \mathbb{E}(- \log_2{\widehat{p}_c}) - H_{\infty}(S) \nonumber \\ 
	    & > - \log_2{\mathbb{E}(\widehat{p}_c)} - H_{\infty}(S) \label{eq:bias_lrs_1} \\
	    & = H_2(S) - H_\infty(S)  \\
	    & \ge 0  \label{eq:bias_lrs_2}
	\end{align}
	where \eqref{eq:bias_lrs_1} follows from Jensen's inequality. Since $-\log x$ is strictly convex and $\widehat{p}_c$ is not constant for non-deterministic sources \eqref{eq:bias_lrs_1} holds. Also, \eqref{eq:bias_lrs_2} follows from Remark~\ref{rem:Renyi}. Hence, $b_{\text{LRS}}(S) > 0$.  
	\end{IEEEproof}
	
	As shown in Fig.~\ref{fig:lrs_estimated}, $H_2(S) - H_\infty(S)$ can be large for BMS with $p < 0.5$. Hence, the LRS estimator suffers from the severe overestimation problem.
	
	The bias of the proposed estimator is given by
	\begin{equation}
	    b_{\text{proposed}}(S) = \mathbb{E}(- \log_2{\widehat{\theta}}) - H_{\infty}(S) > \log_2{\frac{\theta}{\mathbb{E}(\widehat{\theta})}}, \label{eq:bias_proposed} 
	\end{equation}
	where $\log_2{\frac{\theta}{\mathbb{E}(\widehat{\theta})}} \le 0$ since $\widehat{\theta}$ is an estimate of the upper bound on $\theta$ (see Theorem~\ref{thm:theta_ub}). If $\mathbb{E}(- \log_2{\widehat{\theta}}) \simeq - \log_2{\mathbb{E}(\widehat{\theta})}$, then the proposed estimator would be an \emph{underestimated} estimator. 
	
	Since NIST SP 800-90B conservatively estimates the min-entropy, the proposed estimator is better aligned with NIST SP 800-90B than the LRS estimator. Further, we will show that the proposed estimator is unbiased for binary sources (see Corollary~\ref{cor:binary_unbiased} and Remark~\ref{rem:binary_unbiased}).   

    We characterize the bias $b_{\text{proposed}}(S)$ by the \emph{sharp}\footnote{The term ``sharp bound'' means that there exists a distribution that achieves this bound with equality.} lower and upper bounds on $\theta$ for a given collision probability $p_c$. The sharp upper bound on $\theta$ is given in Theorem~\ref{thm:theta_ub}. We derive the sharp lower bound on $\theta$ by using the inverted near-uniform distribution. In~\cite{Hagerty2012entropy}, the inverted near-uniform distribution is defined as $\vect{p}_{\text{INU}}(\psi)= (p_1, \ldots, p_k)$ where
    \begin{equation} \label{eq:inverted_nearuniform}
	    p_i = 
		\begin{cases}
			\psi, & \text{if}~i \in \left\{1, \ldots, \left\lfloor \frac{1}{\psi} \right\rfloor \right\}; \\
			1 - \left\lfloor \frac{1}{\psi} \right\rfloor \psi, & \text{if}~i = \left\lfloor \frac{1}{\psi} \right\rfloor + 1 ; \\
			0,& \text{otherwise}.
		\end{cases}
	\end{equation}
	Note that $\psi = \max \{\vect{p}_{\text{INU}}(\psi)\}$.

	\begin{lemma} \label{lem:round_equal}
		For $\frac{1}{n+1}<\psi \leq\frac{1}{n}$ where $n \in \mathbb{N}$, the following relation holds:
		\begin{equation} \label{eq:round_equal}
	       \left\lfloor \frac{1}{\psi} \right\rfloor = \left\lfloor \frac{1}{M_{2}(\vect{p}_{\text{INU}}(\psi))} \right\rfloor = n. 
	    \end{equation}
	\end{lemma}
	\begin{IEEEproof}
	    If $\psi = \frac{1}{n}$, then 
	    \begin{equation} \label{eq:round_integer}
	       M_{2}(\vect{p}_{\text{INU}}(\psi)) = \frac{1}{n}.    
	    \end{equation}
	    Hence, \eqref{eq:round_equal} holds. 
	    
	    If $\frac{1}{n+1} < \psi < \frac{1}{n}$, then $M_{2}(\vect{p}_{\text{INU}}(\psi))$ is an increasing function of $\psi$. It is because $\frac{d M_{2}(\vect{p}_{\text{INU}}(\psi))}{d \psi} = 2n (n + 1) \{\psi  - \frac{1}{n+1} \} > 0$. By \eqref{eq:round_integer}, we obtain $\frac{1}{n+1} < M_{2}(\vect{p}_{\text{INU}}) < \frac{1}{n}$. Then, \eqref{eq:round_equal} holds.
	\end{IEEEproof}
	
	\begin{remark} \label{rem:inverted_near_uniform}
	For an inverted near uniform distribution, Lemma~\ref{lem:round_equal} shows that the collision entropy is close to the min-entropy since $\psi \simeq M_{2}(\vect{p}_{\text{INU}})$. If $\psi = \frac{1}{n}$, then the collision entropy is the same as the min-entropy.  
	\end{remark}
	
    \begin{theorem} \label{thm:collision_entropy_inequality} 
	For any distribution $\vect{p} = (p_1, \ldots, p_k)$ with $n = \left\lfloor \frac{1}{p_c} \right\rfloor$, the following inequalities hold: 
	\begin{equation} \label{eq:theta_lb_ub}
	    \psi \leq \theta \leq \widehat{\theta}
	\end{equation}
	where 
	\begin{align}
	    \psi &= \frac{\sqrt{n \left\{p_c (n+1) - 1 \right\}}+n}{n(n+1)}, \label{eq:psi_eq} \\
	    \widehat{\theta} &= \frac{\sqrt{(k-1)(p_c k -1)}+1}{k} \label{eq:theta_hat_eq}.
	\end{align}
	\end{theorem}
	\begin{IEEEproof}
    Since $\widehat{\theta}$ is derived in Theorem~\ref{thm:theta_ub}, we need to derive only the sharp lower bound $\psi$. The lower bound $\psi$ is achieved with equality by the inverted near-uniform distribution $\vect{p}_{\text{INU}}(\psi)$~\cite{Hagerty2012entropy,Golic1987relationship}. Hence, we need to identify $\vect{p}_{\text{INU}}(\psi)$ satisfying $M_{2}(\vect{p}_{\text{INU}}(\psi)) = p_c$. Suppose that $\frac{1}{n+1}<\psi \le \frac{1}{n}$ where $n \in \mathbb{N}$ (i.e., $\left\lfloor \frac{1}{\psi} \right\rfloor=n$). By \eqref{eq:inverted_nearuniform} and Lemma~\ref{lem:round_equal}, we obtain $M_{2}(\vect{p}_{\text{INU}}(\psi))= n \psi^2+(1 - n \psi)^2 = p_c$, which leads to \eqref{eq:psi_eq}. 
	\end{IEEEproof}    
	
	\begin{remark}
	    For a given collision probability $p_c$, the sharp lower and upper bounds on the min-entropy are given by
	    \begin{equation}
	        -\log_2{\widehat{\theta}} \le H_{\infty}(\vect{p})\le -\log_2{\psi},
	    \end{equation}
	    where $H_{\infty}(\vect{p}) = -\log_2{\theta}$.  
	\end{remark}

	We note that $\psi$ depends only on $p_c$ because $n = \left\lfloor \frac{1}{p_c} \right\rfloor$. On the other hand, $\widehat{\theta}$ depends on $p_c$ and $k$ (alphabet size $|S|$). For given $p_c$ and $k$, we define the \emph{estimation gap} of $\theta$ as
	\begin{equation}
	    g(p_c, k) =  \widehat{\theta} - \psi,
	\end{equation}
	which is the maximum possible bias. The following theorem shows that the estimation gap increases with $k$. 

	\begin{theorem}
	    For non-deterministic sources, the estimation gap $g(p_c, k) =  \widehat{\theta} - \psi$ increases with $k$. 
	\end{theorem}
	\begin{IEEEproof}
	Since $\psi$ does not depend on $k$, we show that $\widehat{\theta}(k)$ is an increasing function of $k$. The  derivative of $\widehat{\theta}(k)$ is given by
	\begin{equation}
	    \frac{d \widehat{\theta}(k)}{d k} = \frac{p_c k + k - 2 -2\sqrt{(k-1)(p_c k -1)}}{2k^2\sqrt{(k-1)(p_c k -1)}}.
	\end{equation}
	
	We can set $p_c > \frac{1}{k}$ because $p_c = \frac{1}{k}$ means that $\vect{p}$ is the uniform distribution, i.e., $\psi = \widehat{\theta} = \frac{1}{k}$ and $H_{\infty}(\vect{p}) = \log_2{k}$. By the arithmetic-geometric mean inequality, 
	\begin{equation}
	    p_c k + k - 2 = (p_c k -1)+(k-1) \geq 2\sqrt{(k-1)(p_c k - 1)}.
	\end{equation}
	Hence, $\frac{d \widehat{\theta}(k)}{d k} > 0$ for $p_c < 1$. Note that $p_c < 1$ for non-deterministic sources. 
	\end{IEEEproof}
	
	\begin{corollary} \label{cor:binary_unbiased}
	    For binary sources with $k = 2$, the estimation gap is zero, i.e., $g(p_c, k = 2) =0$.
	\end{corollary}
	\begin{IEEEproof}
	    For binary sources, it is clear that $\theta \ge \frac{1}{2}$. For $\theta = \frac{1}{2}$, $\vect{p}$ corresponds to the binary uniform distribution, the gap is zero. For $\theta > \frac{1}{2}$, we can set $n = 1$ in \eqref{eq:round_equal} because $\frac{1}{2} < \psi < 1$. By setting $n=1$ and $k=2$, \eqref{eq:psi_eq} and \eqref{eq:theta_hat_eq} are identical, i.e., $\theta = \psi = \widehat{\theta} = \frac{\sqrt{2 p_c - 1} + 1}{2}$. 
    \end{IEEEproof}
	
	\begin{remark}[Unbiasedness] \label{rem:binary_unbiased}
	The proposed estimator is \emph{unbiased} for binary sources. Since most random sources are binary or can be represented by binary sequences, the proposed estimator improves the accuracy of the LRS estimator. 
	\end{remark}

	
	\begin{figure}[!t] 
		\centering
		\subfloat[]{\includegraphics[width=0.45\textwidth]{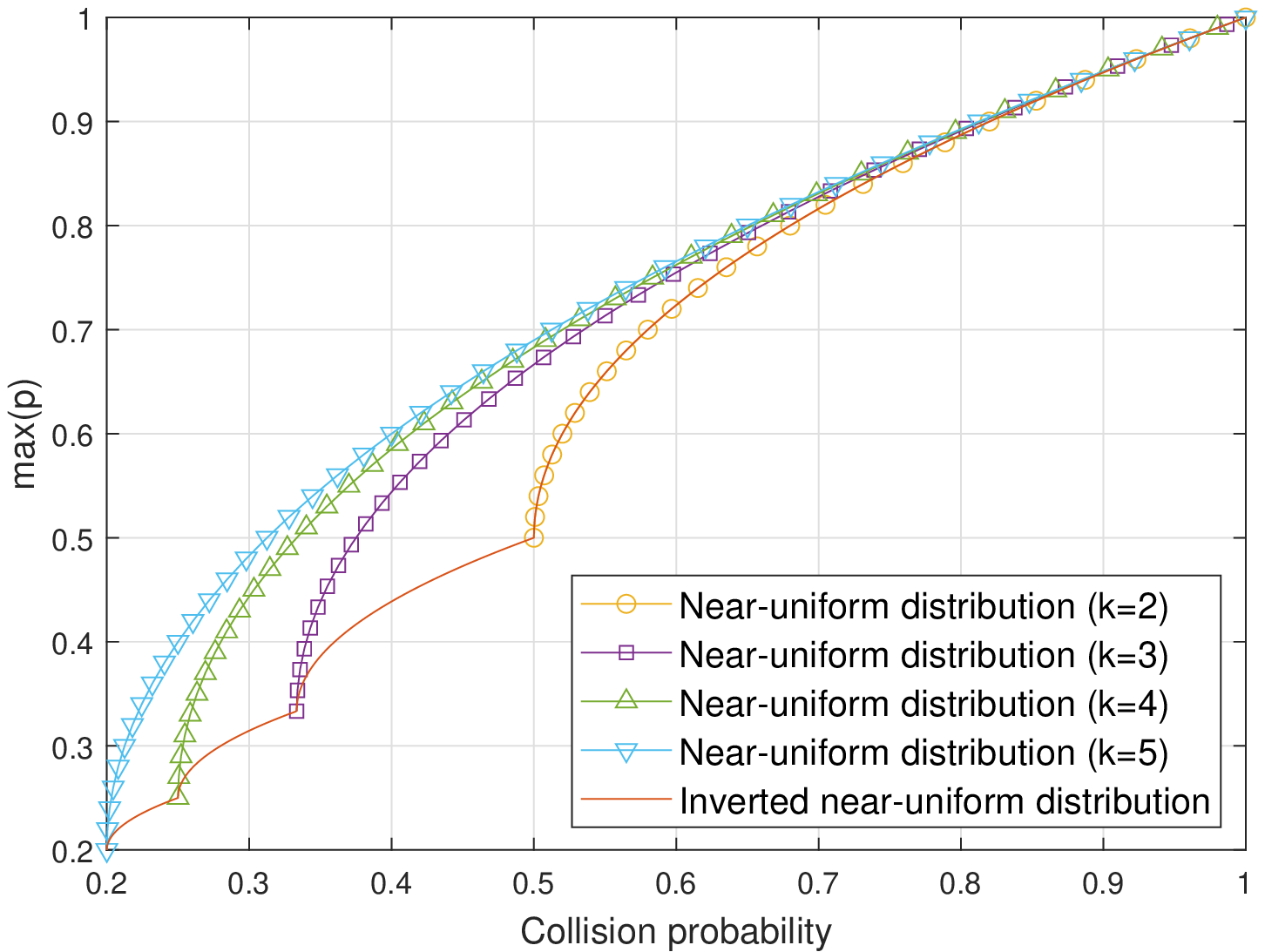}	
			\label{fig:bound_probability}}
		\hfil
		\subfloat[]{\includegraphics[width=0.45\textwidth]{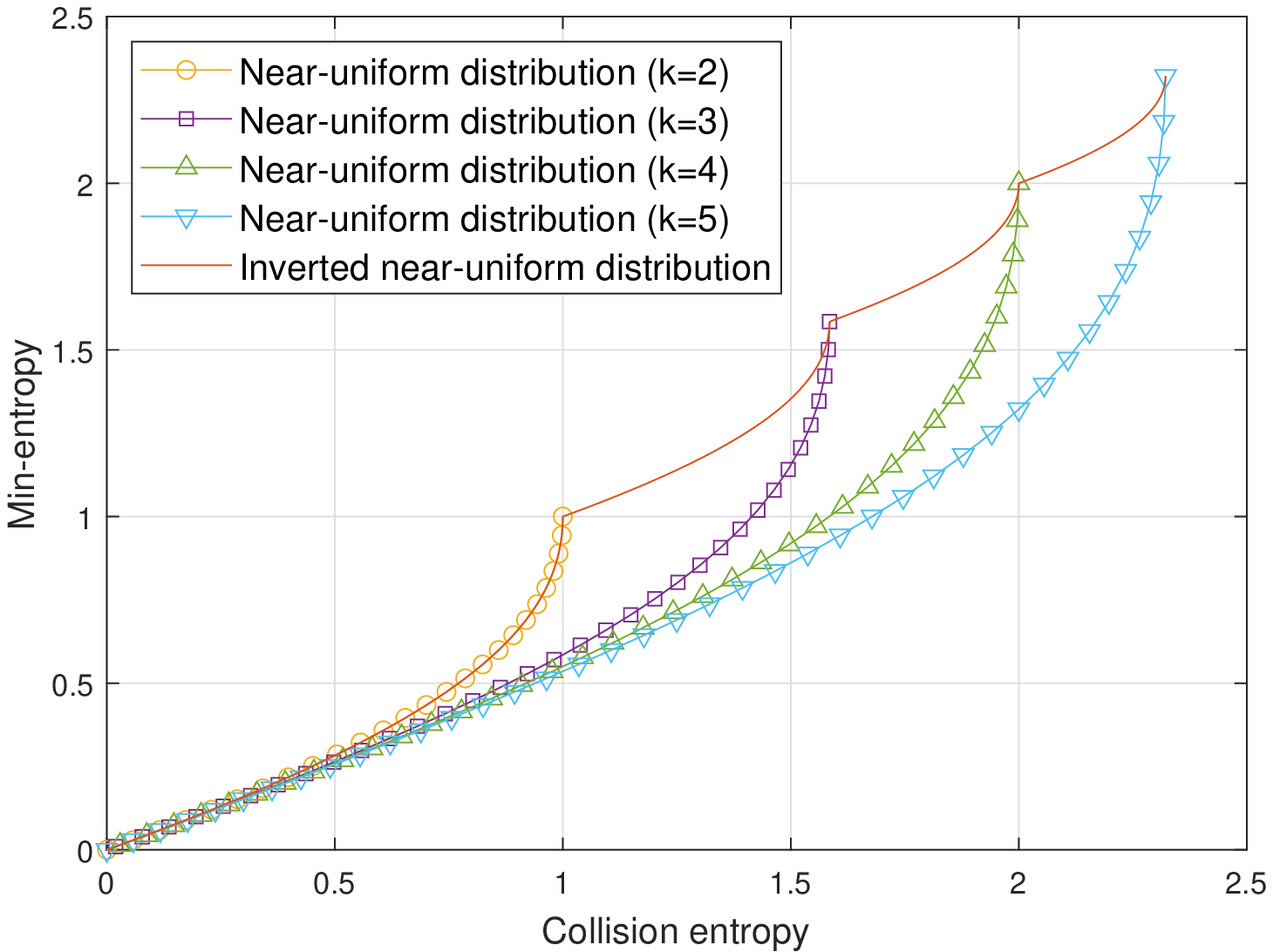}	
			\label{fig:bound_entropy}}
		\caption{The relation of (a) the collision probability $p_c$ and $\theta = \max_{i \in \{1,...,k\}}{p_i}$ and (b) the collision entropy and the min-entropy for $\vect{p}_{\text{NU}}$ and $\vect{p}_{\text{INU}}$.} \label{fig:bound}
	\end{figure}
	
	Fig.~\ref{fig:bound}\subref{fig:bound_probability} shows that the near-uniform distribution and the inverted near-uniform distribution correspond to the upper and lower bounds on $\theta$, respectively. Also, it shows that $\psi = \widehat{\theta}$ for $k = 2$, i.e., the proposed estimator is unbiased for binary souces. Fig.~\ref{fig:bound}\subref{fig:bound_entropy} shows the relation between the collision entropy and the min-entropy where the near-uniform distribution and the inverted near-uniform distribution correspond to the lower and upper bounds on the min-entropy. 
	
    We note that the final min-entropy estimation could not be perfectly unbiased because of the two steps of the original LRS estimator. First, Step 7 of Algorithm~\ref{algo:nist} selects the maximum collision probability among $v - u + 1$ candidates, which lowers the min-entropy estimates. Also, Step 8 of Algorithm~\ref{algo:nist} (or Step 7 of Algorithm~\ref{algo:improved_lrs}) reduces the min-entropy estimates to ensure the confidence level of \unit[99]{\%}. These steps result from the conservative approach of NIST SP 800-90B. These extra-confidence steps are also included in the other estimators of NIST SP 800-90B. 
	
    \section{Generalized LRS Estimator} \label{sec:proposed_generalized}
	
	In this section, we propose a generalized LRS estimator by using the power sum of order $\alpha \ge 2$ instead of the collision probability (the power sum of order $\alpha = 2$). We show that the generalized LRS estimator reduces the variance of estimates as the order $\alpha$ increases beyond 2.
	
	\subsection{Generalized LRS Estimator \label{sec:generalized_lrs}}
	
	\begin{algorithm}[!t] 
		\caption{Generalized LRS estimator} \label{algo:general_lrs}
		\textbf{Input:} Sequence $\vect{s}=(s_1, ..., s_L)$ and an integer $\alpha \ge 2$ \\
		\textbf{Output:} Min-entropy $H_{\infty}(S)$.
		\begin{algorithmic}[1]
			\State Find the smallest $u$ such that the number of occurrences of the most common $u$-tuple in $\vect{s}$ is less than 35.
			\State Find the largest $v$ such that the number of occurrences of the most common $v$-tuple in $\vect{s}$ is at least $\alpha$.
			\For{$w \in \left\{ u, u+1, \ldots,v \right\}$ }
			\State Estimate the $w$-tuple power sum of order $\alpha$: 
			\begin{equation} \label{eq:proposed_collision}
		 	    \widehat{M}_{\alpha,w} := \frac{\sum_i {\binom{C_i}{\alpha}} }{{\binom{l}{\alpha}}},
			\end{equation}
		    	where $C_i$ is the number of occurrences of the $i$th unique $w$-tuple and $l$ is the total number of $w$-tuples.
            \State $\widetilde{M}_{\alpha,w} := \widehat{M}_{\alpha,w}^{\frac{1}{w}}$. 
			\EndFor
            \State $\widetilde{M}_\alpha:=\max\{\widetilde{M}_{\alpha,u}, \ldots,\widetilde{M}_{\alpha,v}\}$.
            \If{$\widetilde{M}_\alpha > \frac{1}{k^{\alpha-1}}$}
			    \State By the bisection method, solve the following equation for $\widehat{\theta}$ $\in \left[\frac{1}{k}, 1\right]$:
			    \begin{equation} \label{eq:proposed_power_sum}
			        \widetilde{M}_\alpha = \widehat{\theta}^{\alpha} +  \frac{(1-\widehat{\theta})^\alpha}{(k-1)^{\alpha-1}}. 
			    \end{equation}

	        \Else
	            \State $\widehat{\theta} := \frac{1}{k}$.
            \EndIf
            \State $\widetilde{\theta}:=\min\left(1,\widehat{\theta}+2.576\sqrt{\frac{\widehat{\theta}(1-\widehat{\theta})}{L-1}}\right)$.
			\State $H_{\infty}(S):=-\log_2\widetilde{\theta}$.
		\end{algorithmic}
	\end{algorithm}	
	
    The generalized LRS estimator is based on 1) the generalized power sum $M_\alpha(\vect{p})$ and 2) the proposed technique in Algorithm~\ref{algo:improved_lrs}. 

    
    The generalized LRS estimator is described in Algorithm~\ref{algo:general_lrs}. First, it estimates the power sum $M_\alpha(\vect{p})$ for a given $\alpha$ by Steps 1--7. Step 2 of Algorithm~\ref{algo:general_lrs} is modified to estimate $M_\alpha(\vect{p})$. Step 4 estimates the $w$-tuple power sum of order $\alpha$ by counting the $\alpha$-wise collisions. Step 5 computes the power sum of order $\alpha$ per sample (to normalize the estimated min-entropy) and Step 7 conservatively chooses the maximum among estimated power sums of $\alpha$, which is denoted by $\widetilde{M}_\alpha$. 
    
    The estimation $\widehat{M}_\alpha(\vect{p})$ by \eqref{eq:proposed_collision} is a key step, which generalizes \eqref{eq:nist_collision} in Algorithm~\ref{algo:nist}. The estimation by \eqref{eq:proposed_collision} is unbiased~\cite{Acharya2015complexity,Bar-Yossef2001sampling}. However, $\widetilde{M}_\alpha$ in Step 7 is an overestimate of $M_\alpha(\vect{p})$, which leads to an underestimate of the min-entropy. We maintain this conservative approach as in the LRS estimator of Algorithm~\ref{algo:nist}. 
    
    Afterward, we estimate $\widehat{\theta}$ from $\widetilde{M}_\alpha$ in Steps 8--12. First, we note that $\widetilde{M}_\alpha \ge \frac{1}{k^{\alpha-1}}$ where the equality is achieved by the uniform distribution. Hence, we set $\widehat{\theta} = \frac{1}{k}$ in Step 11 if $\widetilde{M}_\alpha < \frac{1}{k^{\alpha-1}}$.     If $\widetilde{M}_\alpha > \frac{1}{k^{\alpha-1}}$, then $\widehat{\theta}$ is estimated by \eqref{eq:collision_bound}, which is the sharp upper bound on $\theta$. Especially, $\widehat{\theta}$ is unbiased for binary sources $k = 2$ (see Corollary~\ref{cor:binary_unbiased}). Step 13 ensures the confidence level of \unit[99]{\%} under the Gaussian assumption. Finally, Step 14 estimates the min-entropy from $\widetilde{\theta}$.

    The proposed Algorithm~\ref{algo:general_lrs} improves the bias and reduces the variance compared to the LRS estimator (Algorithm~\ref{algo:nist}). The bias is improved since Algorithm~\ref{algo:general_lrs} estimates the min-entropy whereas the LRS estimator estimates the collision entropy as discussed in Section~\ref{sec:proposed_improved_bias}. The variance can be reduced by using the higher-order power sum instead of the collision probability, which is supported by empirical results in Section~\ref{sec:numerical}. In the following subsection, we provide a theoretical analysis of how the order $\alpha$ affects the variance of estimation. 

    \subsection{Variance of Generalized LRS Estimator \label{sec:var_lrs}}
   	
    In this subsection, we attempt to characterize how the order $\alpha$ affects the variance of $\widehat{\theta}$ calculated by \eqref{eq:proposed_power_sum} in Algorithm~\ref{algo:general_lrs}. 
    
   Suppose that $\widehat{\theta}_\alpha$ and $\widehat{\theta}_{\alpha+1}$ are the estimated $\widehat{\theta}$ in Algorithm~\ref{algo:general_lrs} by using $\widetilde{M}_{\alpha}$ and $\widetilde{M}_{\alpha+1}$, respectively. We characterize the relation between $\alpha$ and $\mathsf{Var}(\widehat{\theta}) = \mathbb{E}(\widehat{\theta}^2) - \mathbb{E}^2(\widehat{\theta})$. We assume that the length-$w$ tuples counted in Algorithm~\ref{algo:general_lrs} are non-overlapping to simplify the analysis. 
   
    \begin{theorem}\label{thm:variance} For a uniformly distributed $\vect{s} = (s_1, \ldots, s_L)$ with a large $L$, the variance ratio's dependence on $\alpha$ is as follows: 
    \begin{equation}
        \xi(\alpha) = \frac{\mathsf{Var}(\widehat{\theta}_{\alpha+1})}{\mathsf{Var}(\widehat{\theta}_\alpha)} \approx \left(\frac{\alpha}{\alpha+1}\right)^4,
    \end{equation}
    where $\approx$ hides multiplicative terms that tend to 1 as $L$ goes to infinity. 
    \end{theorem}
    \begin{IEEEproof} 
        The proof is given in Appendix~\ref{pf:variance_theta}.
    \end{IEEEproof}
    Since $\xi(\alpha) < 1$, $\mathsf{Var}(\widehat{\theta})$ decreases with $\alpha$ for high-entropy sources. The reduction of $\mathsf{Var}(\widehat{\theta}_{\alpha})$ diminishes as $\alpha$ increases.
    
    The range of $w\in\{u, \ldots, v\}$ is an important parameter that affects the variance of $\widehat{\theta}$. It is clear that 
    \begin{equation} \label{eq:v_relation}
        v_\alpha \ge v_{\alpha + 1},   \quad u = u_\alpha = u_{\alpha+1},
    \end{equation}
    where $v_\alpha$ and $v_{\alpha+1}$ are calculated by Step 2 of Algorithm~\ref{algo:general_lrs} for $\alpha$ and $\alpha+1$, respectively. Note that $u$ in Algorithm~\ref{algo:general_lrs} does not depend on $\alpha$. The proof of Theorem~\ref{thm:variance} relies on this relation since the reduction of $v$ leads to the reduction of $\mathsf{Var}(\widehat{\theta})$. 
    
    Theorem~\ref{thm:variance} characterizes $\mathsf{Var}(\widehat{\theta})$ by~\eqref{eq:proposed_power_sum}, i.e., for $\widetilde{M}_\alpha > \frac{1}{k^{\alpha-1}}$. If $\widetilde{M}_\alpha < \frac{1}{k^{\alpha - 1}}$, then Step 11 sets $\widehat{\theta} := \frac{1}{k}$. It is because the power sum of order $\alpha$ cannot be lower than $\frac{1}{k^{\alpha - 1}}$ (attained by the uniform distribution). It is difficult to analyze the probability of $\widetilde{M}_\alpha < \frac{1}{k^{\alpha - 1}}$ due to Step 7 of $\widetilde{M}_\alpha:=\max\{\widetilde{M}_{\alpha,u}, \ldots,\widetilde{M}_{\alpha,v}\}$ in Algorithm~\ref{algo:general_lrs}. 
    
    Although Theorem~\ref{thm:variance} focuses on uniformly distributed sources, the following section empirically supports that $\mathsf{Var}(\widehat{\theta})$ decreases with $\alpha$ even for non-uniformly distributed sources. 
    
    \section{Numerical Results}\label{sec:numerical}

	We evaluate our proposed estimators for simulated and real-world data samples. The empirical results show that the proposed estimator effectively reduces the bias problem of the LRS estimator. 
	
	The following representative samples are considered as in \cite{Kelsey2015predictive,Kim2021efficient}:
	\begin{itemize}
		\item \emph{Binary memoryless source (BMS):} Samples are generated by Bernoulli distribution with $P(S=1) = p$ and $P(S=0) = 1 - p$ (IID);   
		\item \emph{Markov source:} Samples are generated using the first-order Markov model with $P(S_{i+1} = 1|S_i = 0) = P(S_{i+1} = 0| S_i = 1) = p$ (non-IID);
		\item \emph{Near-uniform distribution:} Samples are generated by the near-uniform distribution with $k = 64$ (see \eqref{eq:near_uniform}) (IID); 
		\item \emph{Inverted near-uniform distribution:} Samples are generated by the inverted near-uniform distribution with $k = 64$ (see \eqref{eq:inverted_nearuniform}) (IID). 
	\end{itemize}
	For each of the above sources, one thousand simulated sources were created in each of the above datasets. BMS source and Markov source generate a sequence of $L = 100,000$ bits. The other sources generate a sequence of $L=10,000$ bits and $k=64$.
	
	\begin{figure}[!t]
		\centering 
		\subfloat[]{\includegraphics[width=0.45\textwidth]{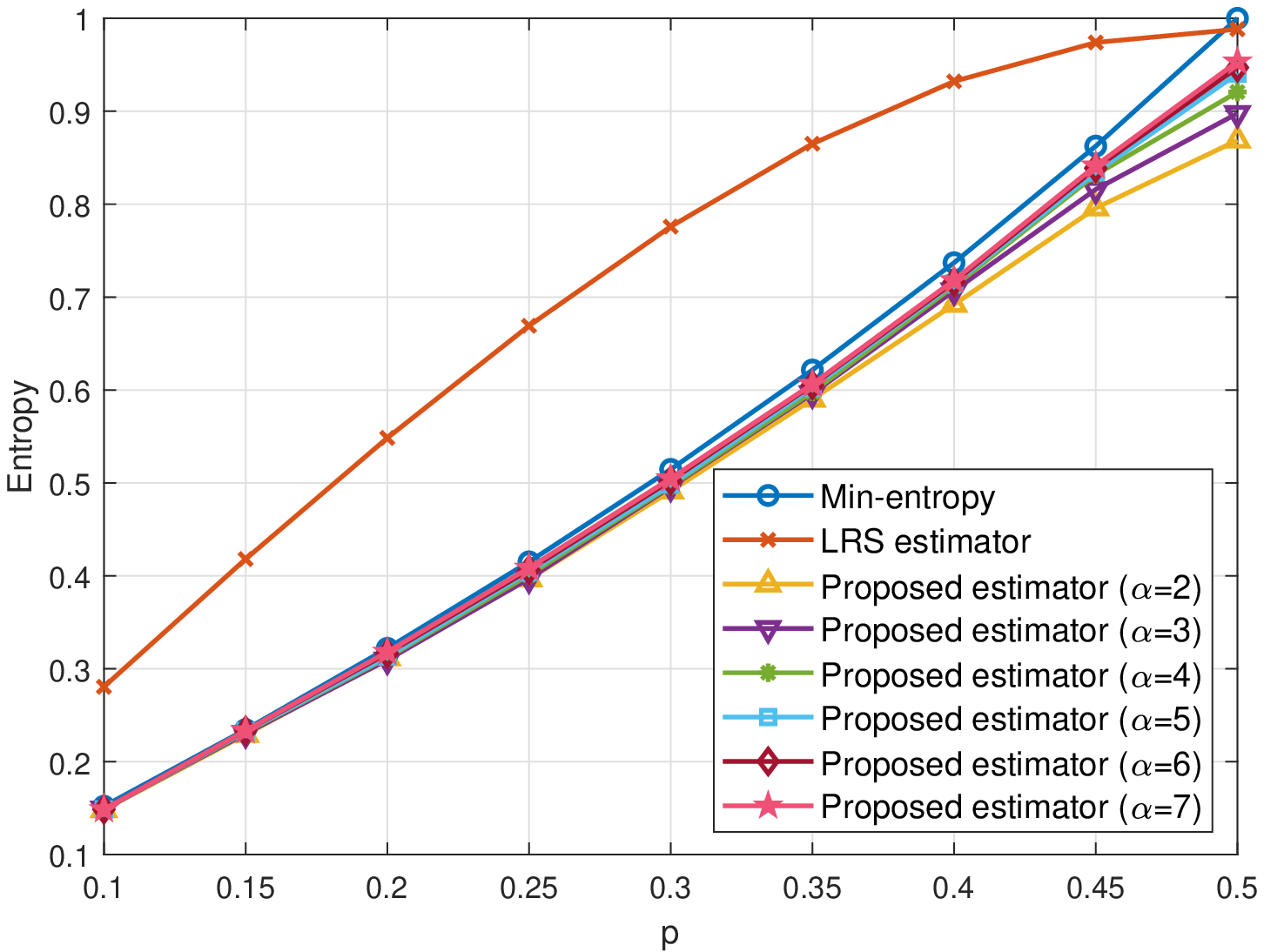}\label{fig:bms_estimate}}
		\hfil
		\subfloat[]{\includegraphics[width=0.45\textwidth]{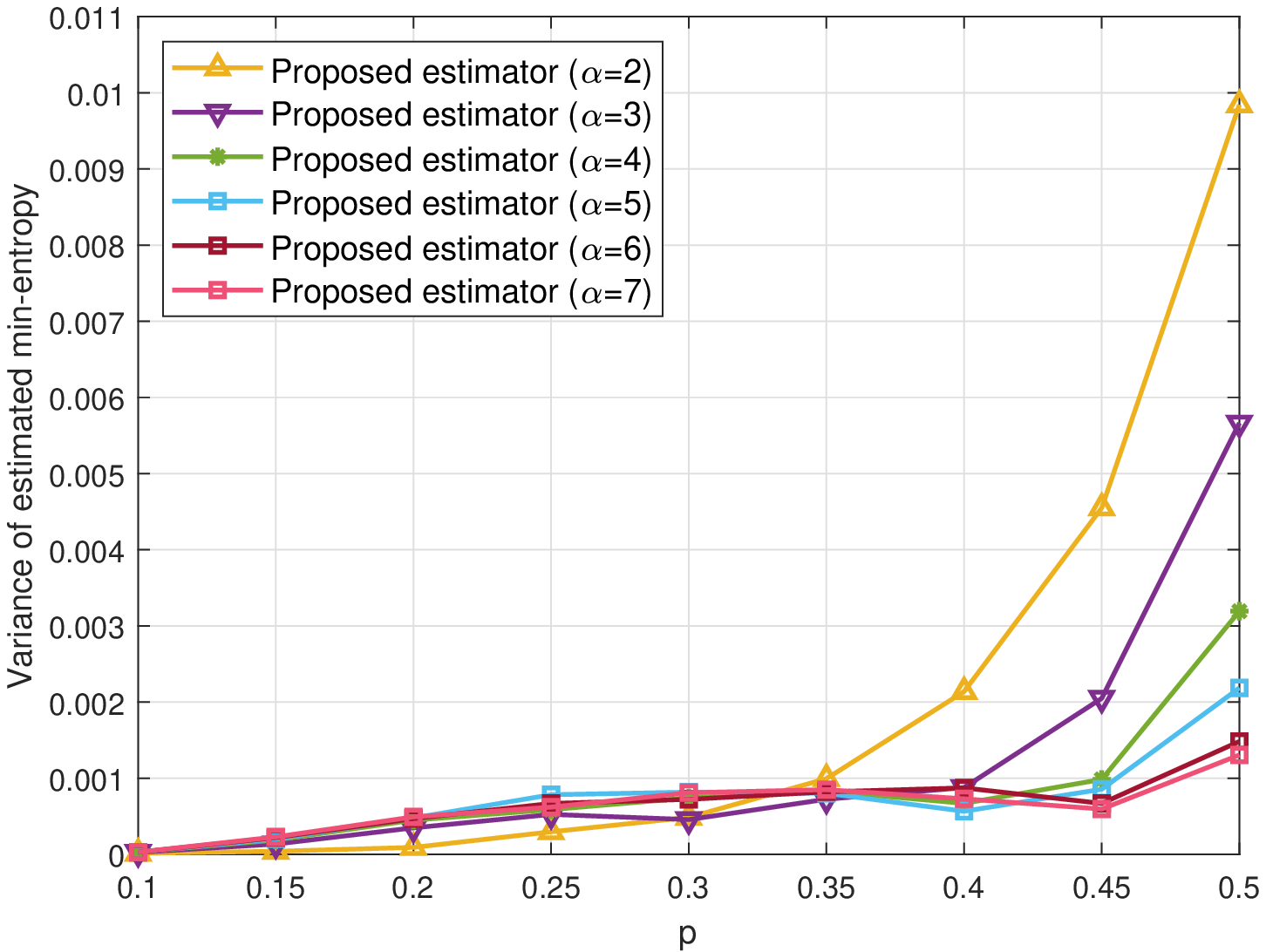}\label{fig:bms_variance}}	
		\caption{(a) Estimated min-entropy and (b) the variance of min-entropy estimates by the proposed generalized LRS estimator for the BMS sources with $p$.}\label{fig:bms} 
	\end{figure}

   \begin{figure}[!t]
    	\centering
    	\subfloat[]{\includegraphics[width=0.45\textwidth]{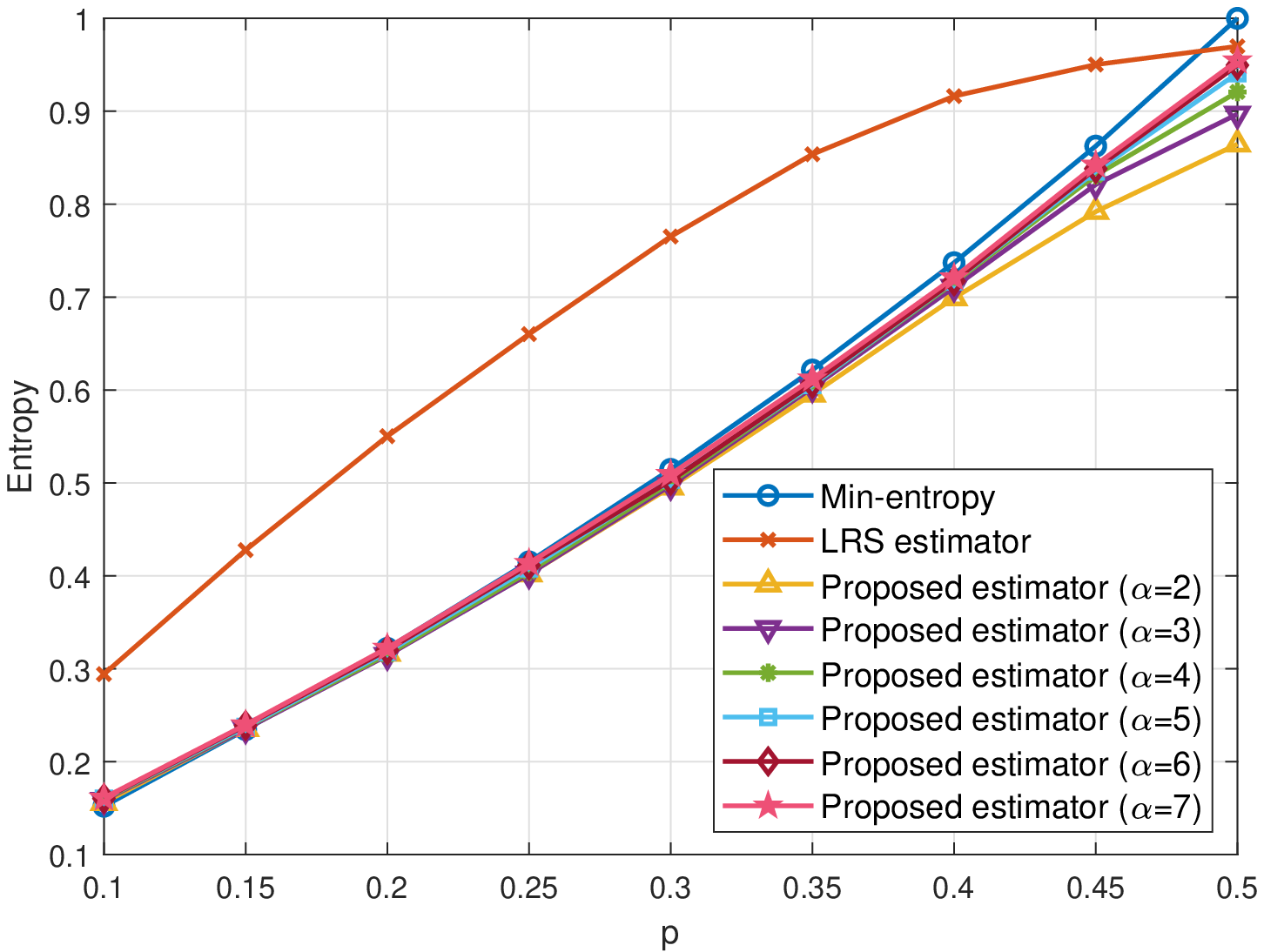}\label{fig:markov_estimate}}
    	\hfil
    	\subfloat[]{\includegraphics[width=0.45\textwidth]{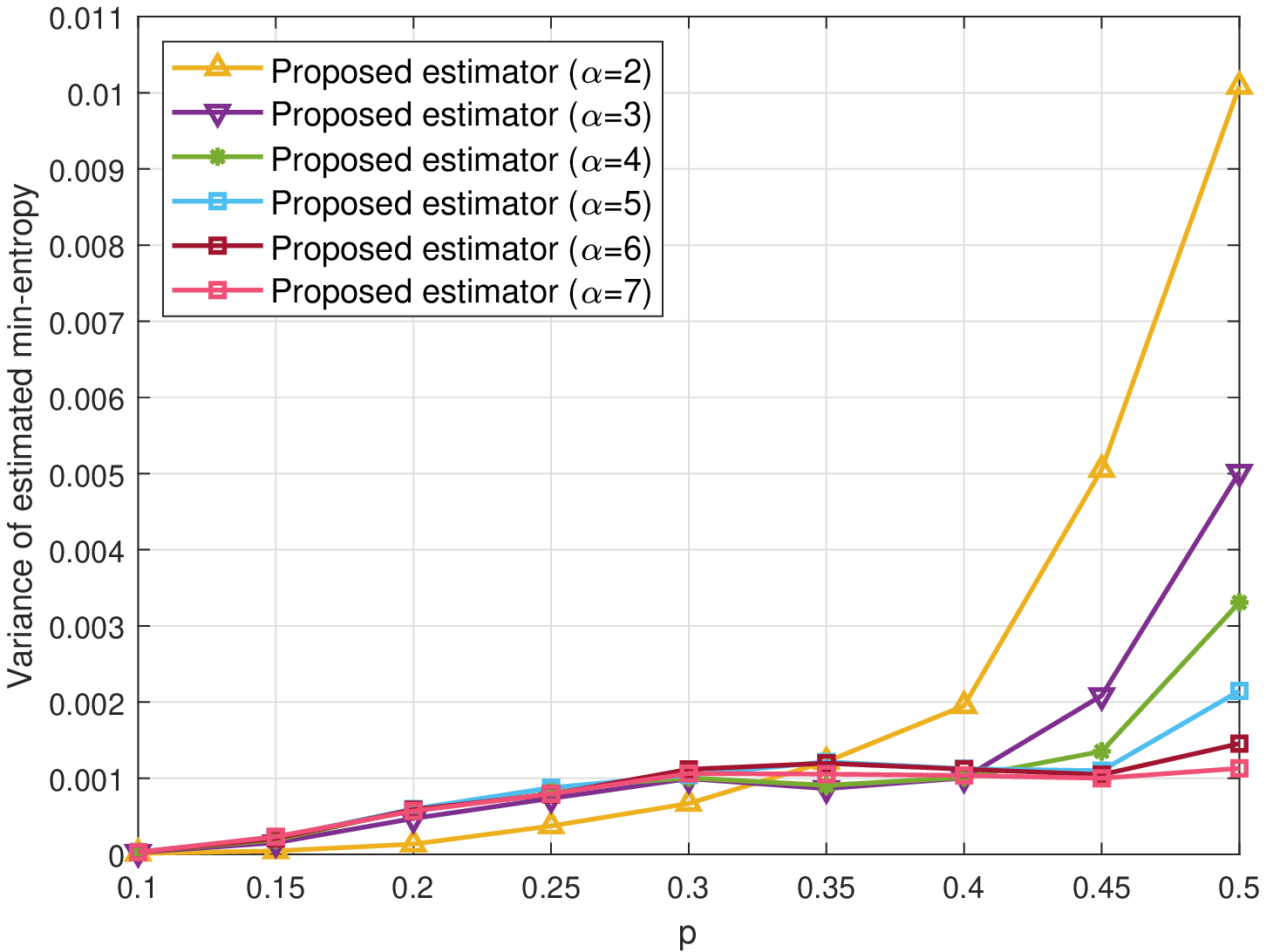}\label{fig:markov_variance}}
    	\caption{(a) Estimated min-entropy and (b) the variance of min-entropy by the proposed generalized LRS estimator for the first-order Markov sources with $p = p(1|0) = p(0|1)$.}\label{fig:markov}
    \end{figure}
 
	Fig.~\ref{fig:bms} compares the min-entropy estimators for BMS as a function of $p$. The theoretical min-entropy and collision entropy are given by $H_{\infty}(S) = - \log_2 \max\{p, 1-p\}$ and $H_2(S) = -\log_2 \{p^2 + (1 - p)^2\}$, respectively. As discussed in Section~\ref{sec:prelim} (viz. Fig.~\ref{fig:lrs_estimated}), the LRS estimator estimates the collision entropy instead of the min-entropy. Since $H_{\infty}(S) \le H_{2}(S)$, the LRS estimator undesirably overestimates the min-entropy. For $p = 0.3$, the bias of the LRS estimator is around 0.28. 
	
	The proposed estimator accurately estimates the min-entropy as shown in Fig.~\ref{fig:bms}\subref{fig:bms_estimate}. As $p \rightarrow 0.5$ (i.e., uniformly distributed sources), the higher $\alpha$ reduces $\mathsf{Var}(\widehat{\theta})$, which supports Theorem~\ref{thm:variance}. We note that the reduction of $\mathsf{Var}(\widehat{\theta})$ diminishes as $\alpha$ increases as shown in~Fig.~\ref{fig:bms}\subref{fig:bms_variance}.

	
    We observe in Fig.~\ref{fig:bms}\subref{fig:bms_estimate} that the higher $\alpha$ slightly improves the bias as $p \rightarrow 0.5$. It is surprising because \eqref{eq:proposed_collision} is unbiased estimator for any $\alpha$. The bias improvement results from Step 11 of Algorithm~\ref{algo:general_lrs}. Since the power sum of order $\alpha$ cannot be lower than $\frac{1}{2^{\alpha-1}}$ for binary sources (i.e., $k=2$), we set $\widehat{\theta} = \frac{1}{2}$ for $\widetilde{M}_\alpha < \frac{1}{2^{\alpha - 1}}$, which leads to $H_{\infty}(S) = 1$. For a BMS source with $p = \frac{1}{2}$, the estimated min entropy would be $1$ (for $\widetilde{M}_\alpha \le \frac{1}{2^{\alpha - 1}}$) or lower than $1$ (for $\widetilde{M}_\alpha > \frac{1}{2^{\alpha - 1}}$). Hence, the increase of $\alpha$ can simultaneously reduce $\mathsf{Var}(\widehat{\theta})$ and improve the bias for high-entropy sources. 
    
    It is worth mentioning that because of the finite sample size $L$, the higher order $\alpha$ reduces the number of valid $C_i$ due to the requirement of $C_i \ge \alpha$ in \eqref{eq:proposed_collision}. Then, \eqref{eq:proposed_collision} could underestimate the power sum of $\alpha$. Given the sample size $L$, an $\alpha$ value should be picked so as to satisfy $C_i\geq \alpha$ for values of $w$ as large as of interest to the randomness tester. In our experiments with $L = 100,000$, we recommend $\alpha\in\{3,4,5,6\}$ by taking into account valid $C_i$ and diminishing variance reduction of $\alpha$.
    
    For the first-order Markov sources, the min-entropy estimators estimate the min-entropy rate. By~\cite{Reched2001renyi,Kamath2016estimation}, the accurate min-entropy rate and the collision entropy rate are given by $H_{\infty}(S) = - \log_2 \max\{p, 1-p\}$ and $H_2(S) = -\log_2 \{p^2 + (1 - p)^2\}$, respectively. Note that the entropy rates of the first-order Markov sources are the same as the entropies of the BMS. 
    
    Fig.~\ref{fig:markov} compares the min-entropy estimators for the first-order Markov sources with parameter $p$. The LRS estimator of NIST SP 800-90B undesirably overestimates the min-entropy of the Markov sources as shown in Fig.~\ref{fig:markov}\subref{fig:markov_estimate}. The proposed estimator effectively improves the accuracy of min-entropy estimates. As in Fig.~\ref{fig:bms}, the generalized estimator improves not only the variance of estimates but also the bias as $p \rightarrow 0.5$. Note that the improvement of $\mathsf{Var}(\widehat{\theta})$ diminishes as $\alpha$ increases as shown in~Fig.~\ref{fig:markov}\subref{fig:markov_variance}. 
    
	\begin{figure}[!t]
		\centering 
		\subfloat[]{\includegraphics[width=0.45\textwidth]{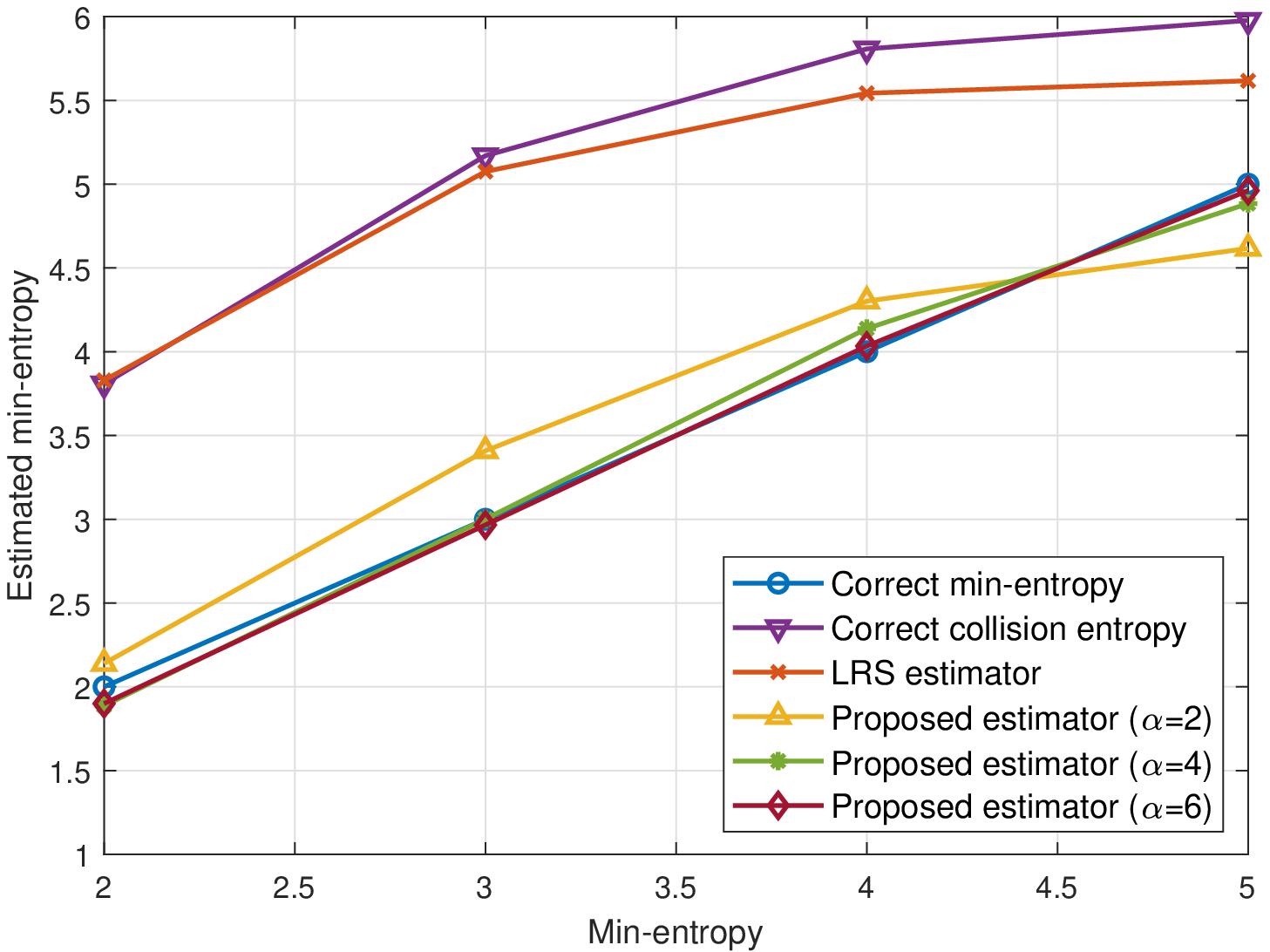}\label{fig:near_uniform}}
		\hfil
		\subfloat[]{\includegraphics[width=0.45\textwidth]{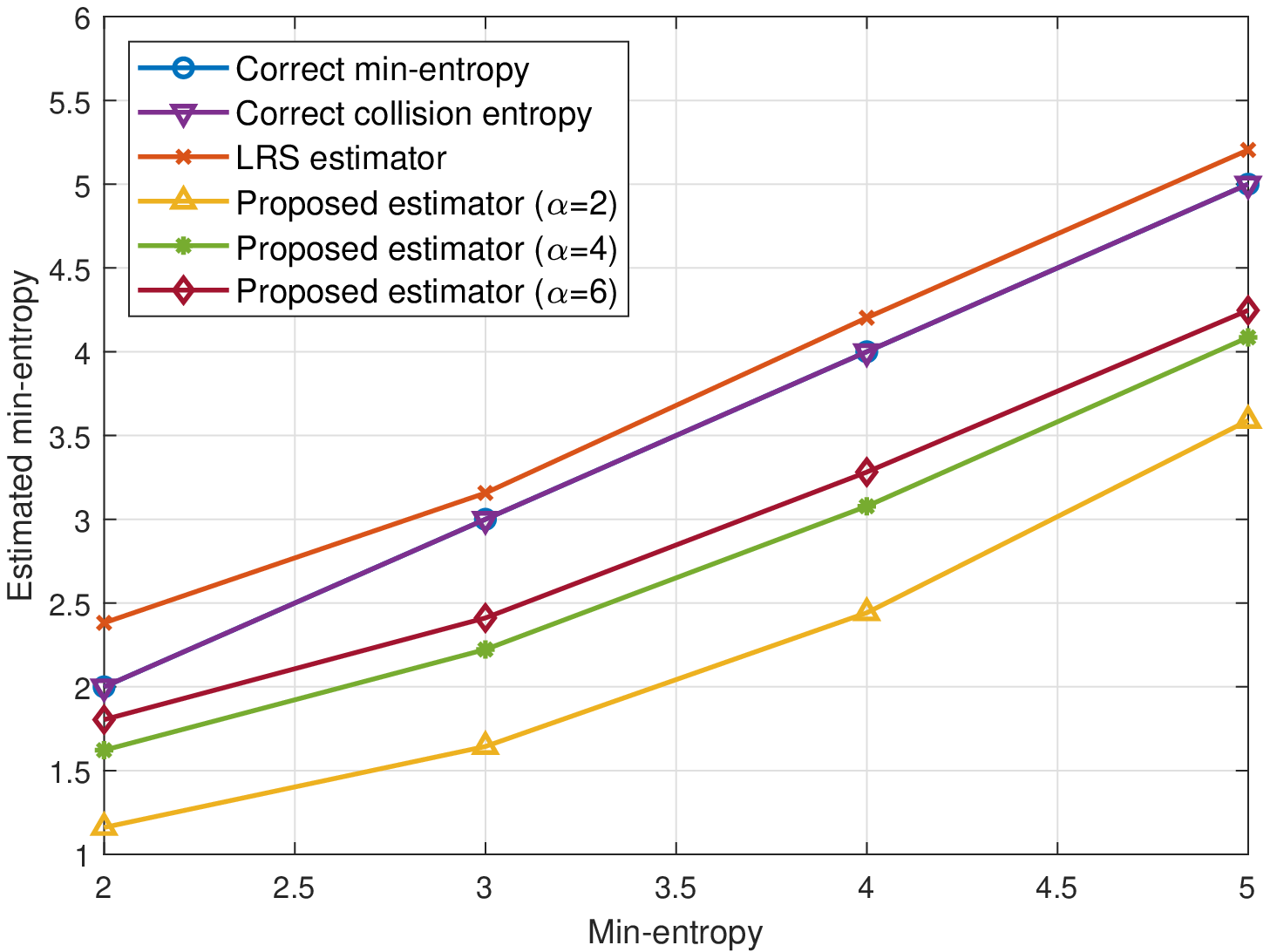}\label{fig:inverted_uniform}}	
		\caption{Comparison of min-entropy estimators for (a) near-uniform distributed sources and (b) inverted near-uniform distributed sources.}\label{fig:non_binary}
		\vspace{-5mm}
	\end{figure}	
	
	
	Fig.~\ref{fig:non_binary} compares the min-entropy estimators for near-uniform distributed sources and inverted near-uniform distributed sources where the alphabet size is 64, i.e., $k=64$. The sequences generated by these non-binary sources are represented by binary values. Then, we estimate the min-entropies from these binary sequences. 
	
	Fig.~\ref{fig:non_binary}\subref{fig:near_uniform} shows that the LRS estimator undesirably overestimates the min-entropy for near-uniform distributed sources since the gap between the collision entropy and the min-entropy is significant. On the other hand, the proposed estimators estimate the min-entropy accurately. We note that the generalized LRS estimator with $\alpha > 2$ (Algorithm~\ref{algo:general_lrs}) is more accurate than the improved LRS estimator (Algorithm~\ref{algo:improved_lrs}). 
	
	For the inverted near-uniform distributed sources, the LRS estimator is relatively accurate (but still overestimating) since the collision entropy is close to the min-entropy as discussed in Remark~\ref{rem:inverted_near_uniform}. Fig.~\ref{fig:non_binary}\subref{fig:inverted_uniform} also supports Remark~\ref{rem:inverted_near_uniform}. We observe that the LRS estimator is close to the collision entropy (and the min-entropy) although it slightly overestimates the min-entropy. The proposed estimators underestimate the min-entropy, which can be explained by \eqref{eq:theta_est}. Since $\widehat{\theta} > \widehat{p}_c$ in \eqref{eq:theta_est}, we observe that $\widehat{\theta} > \widehat{p}_c \simeq \theta$ for inverted near-uniform distributed sources. Fortunately, the underestimation bias can be reduced by adopting a higher order $\alpha$ as shown in Fig.~\ref{fig:non_binary}\subref{fig:inverted_uniform}. 
	
	It is worth mentioning that the compression estimator of NIST SP 800-90B also suffers from this underestimation problem for inverted near-uniform distributed sources~\cite{Turan2018recommendation,Kim2021efficient}. Compared to the compression estimator, our proposed estimators are much more accurate. For an inverted near-uniform distributed source with $H_\infty(S) = 5$, the estimated value by the proposed estimator with $\alpha = 6$ is around 4.3, which is much better than the compression estimator's value of 1.55 (see~\cite[Fig. 6(b)]{Kim2021efficient}). Since NIST SP 800-90B conservatively selects the minimum estimates among ten different estimators including the compression estimator, the proposed generalized LRS estimator does not degrade the final estimation accuracy even for this exceptional inverted near-uniform distribution.      
	
	\begin{table}
		\renewcommand{\arraystretch}{1.2}
		\caption{Min-entropy Estimates for Real-World Sources}
		\vspace{-2mm}
		\label{tab:real}
		\centering
		\begin{tabular}{|c|c|c|c|c|}	\hline
			 & $\alpha=2$ & $\alpha=4$ & $\alpha=6$   \\ \hline \hline
			{RANDOM.ORG}   & 0.8889 & 0.9549 & 0.9572   \\ \hline
			{Ubld.it}  & 0.8277 & 0.8598 &  0.8941  \\ \hline
			{LKRNG}  & 0.9364 & 0.9843 &  0.9844 \\ \hline		  						
		\end{tabular}
		\vspace{-2mm}
	\end{table} 
	
    We also evaluate min-entropy estimates using random number generators deployed in the real-world as in~\cite{Kelsey2015predictive,Kim2021efficient}. The true min-entropies for these sources are unknown even though they are believed to be high-entropy sources. We evaluate RANDOM.ORG, Ubld.it, and Linux kernel random number generator (LKRNG). RANDOM.ORG~\cite{Random} is a service that provides random numbers based on atmospheric noise and Ubld.it generates random numbers by a TrueRNG device by~\cite{Ubld}. The min-entropy estimates of the real-world sources are presented in Table~\ref{tab:real}. We observe that the generalized LRS estimator improves the accuracy of min-entropy estimates by assuming that these real-world sources are high-entropy sources. 
    
    We observe that the LRS estimator suffers from significant overestimation problem for most cases of BMS, Markov sources, and near-uniform distributed sources. Hence, the proposed estimator would be an appealing alternative to the original LRS estimator. 

	\section{Conclusion}\label{sec:conclusion}
	
	We proposed accurate min-entropy estimators to resolve the overestimation problem of the LRS estimator. Although the proposed estimator (improved LRS estimator) relies on the estimated collision probability as in the LRS estimator, it effectively reduces the bias by leveraging the relation between the collision entropy and the min-entropy. Furthermore, we proposed the generalized LRS estimator by parameterizing $\alpha$ instead of setting $\alpha = 2$. It was shown that the generalized LRS estimator can improve the bias and variance of min-entropy estimates.  
	\appendices
	
    \section{Analysis on $v$} \label{pf:w_range}

    In this appendix, we analyze $v$ (i.e., the maximum value of $w$), which is used in Algorithms~\ref{algo:nist} and~\ref{algo:general_lrs}. We denote the number of $\alpha$-wise collisions as $D_{\alpha,w}$ for the $w$-tuples in Step 4 of Algorithm~\ref{algo:general_lrs}, which is given by
    \begin{equation} \label{eq:collision_count}
        D_{\alpha,w} = \sum_{i=1}^{k^w}{\binom{C_i}{\alpha}},
    \end{equation}
    where $C_i$ is the number of occurrences of the $i$th $w$-tuple. We suppose that $\binom{C_i}{\alpha} = 0$ if $C_i < \alpha$. Note that \eqref{eq:collision_count} is the same as the numerator of \eqref{eq:proposed_collision}. 
    
    

    The following lemma shows the relation between the tuple size $w$ and the number of $\alpha$-wise collisions \eqref{eq:collision_count}. 
    
   	\begin{lemma}\label{thm:relation_D}
   	For a large $L$, 
   	$\mathbb{E}(D_{\alpha,w+1}) \approx M_{\alpha} (\vect{p}) \cdot \mathbb{E}(D_{\alpha,w})$. 
   	\begin{IEEEproof}
   	    Denote $\{a_1,\ldots,a_{k^w}\}$ as the alphabet of $w$-tuples and $\{c_1,\ldots,c_{k^w}\}$ as the number of occurrences of each $w$-tuple in $\vect{s}$. For a $w$-tuple element $a_j$ for $j \in \{1, \ldots, k^w\}$, we can represent $a_j$ as $(a_{j, 1},\ldots,a_{j,w})$ where $a_{j,i} \in \{x_1, \ldots, x_k\}$ for $i \in \{1, \ldots, w\}$. 
   	    
   	    For each $w$-tuple $a_j = (a_{j, 1},\ldots,a_{j,w})$, there are $k$ different ways to add a symbol and obtain a $(w+1)$-tuple. The expected numbers of occurrences with $a_j$ as prefix are $\{c_j \cdot p_1, c_j \cdot p_2, \ldots, c_j \cdot p_k \}$. Hence, the expected number of $\alpha$-wise collisions for $(w+1)$-tuples is given by
   	    \begin{align}
   	        \mathbb{E}(D_{\alpha,w+1})\ & = \sum_{j=1}^{k^w}\sum_{i=1}^{k}\binom{c_j p_i}{\alpha} \nonumber \\
   	        & \approx \sum_{j=1}^{k^w}\sum_{i=1}^{k}\frac{c_j^{\alpha}p_i^{\alpha}}{\alpha!}\approx \sum_{j=1}^{k^w} \frac{c_j^{\alpha}M_{\alpha}(\vect{p})}{\alpha!} \label{eq:relation_D_1} \\
   	        &\approx \sum_{j=1}^{k^w} \binom{c_j}{\alpha} \cdot M_{\alpha} (\vect{p}) \label{eq:relation_D_2} \\
   	        & = M_{\alpha}(\vect{p}) \cdot \mathbb{E}(D_{\alpha,w}), \label{eq:relation_D_3}
   	    \end{align}
   	    where \eqref{eq:relation_D_1} follows from $\binom{c_j p_i}{\alpha} \approx \frac{c_j^\alpha p_i^\alpha}{\alpha !}$ for $c_j p_i \gg \alpha$ (i.e., for a large $L$) and Definition~\ref{def:power_sum}. If a $c_j p_i$ is not much greater than $\alpha$, then it can be neglected. Also, \eqref{eq:relation_D_2} follows from $\binom{c_j}{\alpha} \approx \frac{c_j^\alpha}{\alpha !}$. Finally, \eqref{eq:relation_D_3} follows from \eqref{eq:collision_count}. 
   	\end{IEEEproof}
   	\end{lemma}
   	
    For the proof of Theorem~\ref{thm:variance}, we will take the value of $v$ to be $\overline{v}$, which is defined to be the tuple length at which the distribution attains \emph{in expectation} the cutoff property of having at least one tuple occurring at least $\alpha$ times in the sequence (see Step 2 in Algorithm~\ref{algo:general_lrs}).  
    
  	\begin{lemma}\label{thm:determined_v}
   	For a large $L$, $\overline{v}  \approx \log_{\frac{1}{M_{\alpha}}} \binom{l}{\alpha}$.
   	\begin{IEEEproof}
   	    By the definition of $\overline{v}$ and \eqref{eq:collision_count}, $\overline{v}$ is the largest value such that $\mathbb{E}(D_{\alpha,v}) \ge 1$. Hence, $\mathbb{E}(D_{\alpha,1})\cdot(M_{\alpha})^{\overline{v} - 1} \ge 1$ and $\mathbb{E}(D_{\alpha,1})\cdot(M_{\alpha})^{\overline{v}} < 1$ by Lemma~\ref{thm:relation_D}. Then, we can obtain
   	    \begin{align}
   	       &\log_{\frac{1}{M_{\alpha}}}\mathbb{E}(D_{\alpha,1}) < \overline{v} \le  \log_{\frac{1}{M_{\alpha}}}\mathbb{E}(D_{\alpha,1}) + 1 \\
   	       &\log_{\frac{1}{M_{\alpha}}}\binom{l}{\alpha}M_{\alpha} < \overline{v} \le  \log_{\frac{1}{M_{\alpha}}}\binom{l}{\alpha}M_{\alpha} + 1 \\
   	       &\log_{\frac{1}{M_{\alpha}}}\binom{l}{\alpha}-1 < \overline{v} \le \log_{\frac{1}{M_{\alpha}}}\binom{l}{\alpha}. 
   	    \end{align}
   	      
   	    For $l \gg \alpha $, $\overline{v} \approx \log_{\frac{1}{M_{\alpha}}}\binom{l}{\alpha}$. 
   	\end{IEEEproof}
   	\end{lemma}
   	
    \begin{lemma}\label{thm:v_ratio}
   	For a uniformly distributed $\vect{s} = (s_1, \ldots, s_L)$ with a large $L$,
   	\begin{equation}
   	    \frac{\overline{v}_{\alpha+1}}{\overline{v}_\alpha} \approx \frac{\alpha^2 - 1}{\alpha^2}, 
   	\end{equation}
   	which is less than one. Note that $\overline{v}_\alpha$ denotes the $\overline{v}$ with order $\alpha$. %
   	\begin{IEEEproof}
   	We can obtain
	\begin{align}
	    \frac{\overline{v}_{\alpha+1}}{\overline{v}_\alpha}
	    & = \frac{\log_{k}\binom{l_{\alpha+1}}{\alpha+1}}{\alpha} \label{eq:ratio_v_0} \cdot \frac{\alpha-1}{\log_{k}\binom{l_{\alpha}}{\alpha}} \\
	    & = \frac{\alpha-1}{\alpha} \cdot \frac{\ln\binom{l_{\alpha+1}}{\alpha+1}}{\ln\binom{l_{\alpha}}{\alpha}} \label{eq:ratio_v_1} \\
	    & \approx \frac{\alpha-1}{\alpha} \cdot \frac{\ln L^{\alpha+1} - \ln \overline{v}_{\alpha+1}^{\alpha+1}-\ln (\alpha+1)!}{\ln L^{\alpha}-\ln \overline{v}_{\alpha}^{\alpha}-\ln\alpha!} \label{eq:ratio_v_2} \\
        & \approx \frac{\alpha-1}{\alpha} \cdot \frac{\ln L^{\alpha+1}}{\ln L^\alpha} \label{eq:ratio_v_3} \\
        & = \frac{(\alpha-1)(\alpha+1)}{\alpha^2}, \label{eq:ratio_v_4}
	\end{align}
	where \eqref{eq:ratio_v_0} follows from Lemma~\ref{thm:determined_v} and $M_\alpha=k^{-(\alpha-1)}$ for a uniformly distributed source. For a large $L$, \eqref{eq:ratio_v_2} follows from $\binom{l_\alpha}{\alpha} \approx \frac{l_\alpha^\alpha}{\alpha !}$ and $l_\alpha = \left \lfloor \frac{L}{\overline{v}_\alpha} \right\rfloor \approx \frac{L}{\overline{v}_\alpha}$. Also, \eqref{eq:ratio_v_3} follows from $L^\alpha \gg \overline{v}_\alpha^\alpha$ and $L^\alpha \gg \alpha!$ for a large $L$. 
   	\end{IEEEproof}
   	\end{lemma}	    
	
	\section{Proof of Theorem~\ref{thm:variance}} \label{pf:variance_theta}
	
    For every subset $I \subseteq \{1,\ldots,l = \left \lfloor\frac{L}{w} \right \rfloor\}$ of size $\alpha$, we define $X_I$ be a 0-1 random variable that gets the value 1 iff all the values ${x_i}$ are the same (i.e., $I$ forms a $\alpha$-wise collision). By \eqref{eq:collision_count}, it is clear that
    \begin{equation}
        D_{\alpha,w} = \sum_{|I|=\alpha}{X_I}
    \end{equation} 
    and 
    \begin{equation}
        \mathbb{E}(X_I) = M_{\alpha,w},
    \end{equation} where $M_{\alpha,w}$ is the $w$-tuple power sum of order $\alpha$. Also, we set $\overline{X}_I = X_I - M_{\alpha,w}$ as in~\cite{Bar-Yossef2001sampling}.
    
    For two subsets $I$ and $J$ such that $|I| = |J| = \alpha$, $\mathbb{E}(\overline{X}_I \cdot \overline{X}_J) = \mathbb{E}(\overline{X}_I) \cdot \mathbb{E}(\overline{X}_J) = 0$ if $I \cap J = \emptyset$. If $I \cap J \ne \emptyset$, then $X_I \cdot X_J$ is a 0-1 random variable that gets the value 1 iff all the values in $I \cup J$ are the same. Hence, 
    \begin{equation}
        \mathbb E(\overline{X}_I\cdot\overline{X}_J) = M_{\alpha+t,w} - M_{\alpha,w}^2
    \end{equation}
    if $|I \cup J| = \alpha + t < 2\alpha$~\cite{Bar-Yossef2001sampling}. Since $M_{\alpha,w} = \frac{1}{k^{w(\alpha - 1)}}$ for a uniformly distributed source, we obtain
    \begin{equation} \label{eq:avg_xixj_bar}
        \mathbb E(\overline{X}_I\cdot\overline{X}_J) = \frac{1}{k^{w(\alpha+t-1)}}-\frac{1}{k^{2w(\alpha-1)}}.
    \end{equation}
    
    The variance of $D_{\alpha,w}$ is given by
    \begin{align}
        & \mathsf{Var}(D_{\alpha,w}) \nonumber \\
        & = \sum_{t=0}^{\alpha-1} \sum_{| I\cup J| = \alpha+t} {\mathbb{E}(\overline{X}_I\cdot\overline{X}_J)} \label{eq:var_d_1} \\
        & = \sum_{t=0}^{\alpha-1} \sum_{| I\cup J| = \alpha+t} \left(\frac{1}{k^{w(\alpha+t-1)}} - \frac{1}{k^{2w(\alpha-1)}}\right) \label{eq:var_d_2} \\
        & = \sum_{t=0}^{\alpha-1}{\binom{l}{\alpha}}{\binom{l-\alpha}{t}}{\binom{\alpha}{t}}\left(\frac{1}{k^{w(\alpha+t-1)}}-\frac{1}{k^{2w(\alpha-1)}}\right)  \\
        & \approx \frac{1}{k^{w(\alpha-1)}} {\binom{l}{\alpha}} \sum_{t=0}^{\alpha-2}{\binom{l}{t}}{\binom{\alpha}{t}}\left(\frac{1}{k^{wt}}-\frac{1}{k^{w(\alpha-1)}}\right), \label{eq:var_d_4} 
    \end{align}
    where \eqref{eq:var_d_2} follows from \eqref{eq:avg_xixj_bar}. Also, \eqref{eq:var_d_4} follows from $\binom{l-\alpha}{t} \approx \binom{l}{t}$ for $l \gg \alpha$ and $\frac{1}{k^{wt}} - \frac{1}{k^{w(\alpha-1)}} = 0$ for $t = \alpha - 1$. 
    
    By taking into account normalization in Step 5 of Algorithm~\ref{algo:general_lrs}, we obtain
    \begin{align}
	    & \mathsf{Var}(\widetilde{M}_{\alpha,w}) = \mathsf{Var}\left(\widehat{M}_{\alpha,w}^{\frac{1}{w}}\right) \nonumber \\ 
	    & \approx \frac{1}{w^2} \cdot \mathbb E(\widehat{M}_{\alpha,w})^{\frac{2(1-w)}{w}}\cdot \mathsf{Var}(\widehat{M}_{\alpha,w}) 
	    \label{eq:var_norm_1} \\ 
	    & = \frac{1}{w^2} \cdot k^{2(\alpha - 1)(w - 1)} \cdot \frac{\mathsf{Var}(D_{\alpha,w})}{\binom{l}{\alpha}^2}\label{eq:var_norm_2} \\
	    & \approx \frac{k^{(\alpha-1)(w-2)}}{w^2} \cdot \frac{\sum_{t=0}^{\alpha-2}\binom{l}{t}\binom{\alpha}{t} \left({k^{-wt}} - {k^{-w(\alpha-1)}}\right)}{\binom{l}{\alpha}}, \label{eq:var_norm_3} 
	\end{align}    
	where \eqref{eq:var_norm_1} follows from the first-order Taylor approximation (i.e., $\mathsf{Var}(f(x)) \approx f'(\mathbb{E}(x))^2 \cdot \mathsf{Var}(x)$ where $f(x) = x^{\frac{1}{w}}$). Since \eqref{eq:proposed_collision} is an unbiased estimator (i.e., $\mathbb{E}(\widehat{M}_{\alpha,w}) = M_{\alpha,w} = k^{-w(\alpha - 1)}$), \eqref{eq:var_norm_2} holds. Finally, \eqref{eq:var_norm_3} follows from \eqref{eq:var_d_4}. 

    Now we show that $\mathsf{Var}(\widetilde{M}_{\alpha,w}) \le \mathsf{Var}(\widetilde{M}_{\alpha,w+1})$ for $w \ge 3$. For each term of \eqref{eq:var_norm_3},
	\begin{align}
	    & \frac{k^{(\alpha-1)(w-2)}}{w^2} \cdot \frac{\binom{l}{t}\binom{\alpha}{t}\left({k^{-wt}} - {k^{-w(\alpha-1)}}\right)}{\binom{l}{\alpha}} \nonumber \\
	    & < \frac{k^{(\alpha-1)(w-2)}}{(w+1)^2} \cdot \frac{k^{\alpha-1}}{k^t} \cdot \frac{\binom{l}{t}\binom{\alpha}{t}\left({k^{-wt}} - {k^{-w(\alpha-1)}}\right)}{\binom{l}{\alpha}} \label{eq:var_w_1} \\
	    & =\frac{k^{(\alpha-1)(w-1)}}{(w+1)^2} \cdot \frac{\binom{l}{t}\binom{\alpha}{t}\left({k^{-(w+1)t}} - {k^{- \{w(\alpha-1) + t\}}}\right)}{\binom{l}{\alpha}} \label{eq:var_w_2} \\
	    & < \frac{k^{(\alpha-1)(w-1)}}{(w+1)^2} \cdot \frac{\binom{l}{t} \binom{\alpha}{t} \left({k^{-(w+1)t}} - {k^{- (w+1)(\alpha-1)}}\right)}{\binom{l}{\alpha}}, \label{eq:var_w_3}
	\end{align}
	where \eqref{eq:var_w_1} follows from $\left(\frac{w}{w+1}\right)^2\cdot\frac{k^{\alpha-1}}{k^t} > 1$ for $k \ge 2$ and $w \ge 3$. Also, \eqref{eq:var_w_3} follows from $t < \alpha - 1$. Hence, 
	\begin{equation} \label{eq:var_w_v}
	    \mathsf{Var}(\widetilde{M}_{\alpha,w}) < \mathsf{Var}(\widetilde{M}_{\alpha,w+1})
	\end{equation}
	for $w \ge 3$. 
	
	In Step 7 of Algorithm~\ref{algo:general_lrs}, the maximum among $\{\widetilde{M}_{\alpha,u}, \ldots, \widetilde{M}_{\alpha,v}\}$ is chosen as $\widetilde{M}_\alpha$. It is difficult to characterize which $\widetilde{M}_{\alpha,w}$ for $w \in \{u, \ldots, v\}$ is the maximum value. As a conservative approach, we set $\mathsf{Var}(\widetilde{M}_{\alpha})  \approx  \mathsf{Var}(\widetilde{M}_{\alpha,\overline{v}})$. Then,
	\begin{align} 
	\mathsf{Var}(\widetilde{M}_{\alpha}) & \approx \frac{k^{(\alpha-1)(\overline{v}_\alpha - 2)}}{\overline{v}_\alpha^2} \nonumber \\
	& \quad  \cdot \frac{\sum_{t=0}^{\alpha-2}\binom{l_\alpha}{t}\binom{\alpha}{t} \left({k^{-t \overline{v}_\alpha}} - {k^{-(\alpha-1)\overline{v}_\alpha}}\right)}{\binom{l_\alpha}{\alpha}}, \label{eq:var_v}
	\end{align}
	where we denote $\overline{v} = \overline{v}_\alpha$ and $l = l_\alpha$ since both $\overline{v}$ and $l$ depend on $\alpha$.
	
    By the first-order Taylor approximation, 
	\begin{equation}\label{eq:variance_theta}
  	    \mathsf{Var}(\widehat{\theta}_\alpha) \approx z(\widehat{\theta}_\alpha,\alpha)^2 \cdot \mathsf{Var}(\widetilde{M}_{\alpha}),
 	\end{equation}
 	where $z(\widehat{\theta}_\alpha,\alpha)$ is given by
 	\begin{equation} \label{eq:def_z}
		z(\widehat{\theta}_\alpha, \alpha) = \frac{d \widehat{\theta}_\alpha}{d \widetilde{M}_{\alpha}} = \frac{1}{\alpha \left\{ \widehat{\theta}_{\alpha}^{\alpha - 1} -  \left(\frac{1 - \widehat{\theta}_\alpha}{k-1} \right)^{\alpha - 1} \right\}}, 
	\end{equation}
	which is derived from \eqref{eq:proposed_power_sum}. 
	
	If $\widetilde{M}_\alpha = \frac{1}{k^{\alpha - 1}}$, then $z(\widehat{\theta}_\alpha, \alpha) \rightarrow \infty$. However, Algorithm~\ref{algo:general_lrs} sets $\widehat{\theta}_\alpha = \frac{1}{k}$ for $\widetilde{M}_\alpha \le \frac{1}{k^{\alpha - 1}}$ instead of solving \eqref{eq:proposed_power_sum}. Hence, $z(\widehat{\theta}_\alpha, \alpha)$ should be considered only if $\widetilde{M}_\alpha = \frac{1}{k^{\alpha - 1}} + \delta$ where $0 < \delta \ll k$ for uniformly distributed sources. Then, we can set $\widehat{\theta}_\alpha = \frac{1}{k} + \delta'$ where $0 < \delta' \ll k$. By~\cite[Theorem 4]{Kim2021efficient}, $z(\widehat{\theta}_\alpha, \alpha) \approx \frac{k^{\alpha-3}}{\alpha (\alpha - 1)} \cdot
	    \frac{k-1}{\delta'}$. Then, 
	\begin{equation} \label{eq:z_ratio}
	    \frac{z(\widehat{\theta}_{\alpha+1}, \alpha+1)}{z(\widehat{\theta}_\alpha, \alpha)} \approx \frac{\alpha - 1}{\alpha + 1} \cdot k. 
	\end{equation}
	
	Then, we obtain 
	\begin{align}
	    \xi(\alpha) & = \frac{\mathsf{Var}(\widehat{\theta}_{\alpha+1})}{\mathsf{Var}(\widehat{\theta}_{\alpha})} \nonumber \\
   	    & \approx \frac{z(\widehat{\theta}_{\alpha+1},\alpha+1)^2}{z(\widehat{\theta}_\alpha,\alpha)^2} \cdot \frac{\mathsf{Var}(\widetilde{M}_{\alpha+1})}{ \mathsf{Var}(\widetilde{M}_{\alpha})} \label{eq:xi_s1_1} \\
        & = \left(\frac{\alpha-1}{\alpha+1}\right)^2 \cdot \left( \frac{\overline{v}_{\alpha}}{\overline{v}_{\alpha+1}} \right)^2 \cdot \frac{\binom{l_\alpha}{\alpha}}{\binom{l_{\alpha+1}}{\alpha+1}} \cdot k^{ \left\{ \alpha \overline{v}_{\alpha+1} - (\alpha - 1) \overline{v}_{\alpha} \right\}}
        \nonumber \\
        & \quad \cdot \frac{\sum_{t=0}^{\alpha - 1} \binom{l_{\alpha+1}}{t}\binom{\alpha+1}{t}\left({k^{ - t \overline{v}_{\alpha+1}}} - {k^{-\alpha \overline{v}_{\alpha+1}}}\right)} {\sum_{t=0}^{\alpha-2}\binom{l_{\alpha}}{t}\binom{\alpha}{t}\left( {k^{ - t \overline{v}_\alpha}} - {k^{ - (\alpha-1) \overline{v}_\alpha}}\right)} \label{eq:xi_s1_3}\\
        & \approx \left(\frac{\alpha-1}{\alpha+1}\right)^2\cdot
        \left( \frac{\overline{v}_{\alpha }}{\overline{v}_{\alpha+1}} \right)^2
        \nonumber \\
        & \quad \cdot \frac{\sum_{t=0}^{\alpha - 1} \binom{l_{\alpha+1}}{t}\binom{\alpha+1}{t}\left({k^{ - t \overline{v}_{\alpha+1}}} - {k^{-\alpha \overline{v}_{\alpha+1}}}\right)} {\sum_{t=0}^{\alpha-2}\binom{l_{\alpha}}{t}\binom{\alpha}{t}\left( {k^{ - t \overline{v}_\alpha}} - {k^{ - (\alpha-1) \overline{v}_\alpha}}\right)}, \label{eq:xi_s1_4}
	\end{align}
	where \eqref{eq:xi_s1_1} follows from \eqref{eq:variance_theta}. Also, \eqref{eq:xi_s1_3} follows from \eqref{eq:var_v} and \eqref{eq:z_ratio}. By Lemma~\ref{thm:determined_v} (see Appendix~\ref{pf:w_range}) and $M_{\alpha}=\frac{1}{k^{\alpha-1}}$, we obtain $ \overline{v}_\alpha \approx \log_{k^{\alpha - 1}}\binom{l_\alpha}{\alpha} = \frac{1}{\alpha - 1} \log_k{\binom{l_\alpha}{\alpha}}$, which leads to
    \begin{equation}\label{eq:power_k_v}
        k^{\overline{v}_\alpha} \approx \binom{l_\alpha}{\alpha}^{\frac{1}{\alpha-1}}.
    \end{equation} 
    Then, \eqref{eq:xi_s1_4} follows from $k^{ \left\{ \alpha \overline{v}_{\alpha+1} - (\alpha - 1) \overline{v}_{\alpha} \right\}} \approx \frac{\binom{l_{\alpha+1}}{\alpha+1}}{\binom{l_\alpha}{\alpha}}$.
	


	Also, we obtain
   	\begin{align}
   	    \xi(\alpha) 
        & \approx \left(\frac{\alpha}{\alpha+1}\right)^4 \nonumber \\
        & \quad \cdot \frac{\sum_{t=0}^{\alpha-1}\binom{l_{\alpha+1}}{t}\binom{\alpha+1}{t}\left({k^{-t \overline{v}_{\alpha+1}}} - {k^{-\alpha \overline{v}_{\alpha+1}}}\right)} {\sum_{t=0}^{\alpha-2}\binom{l_{\alpha}}{t}\binom{\alpha}{t}\left({k^{ - t \overline{v}_\alpha}} - {k^{ - (\alpha-1) \overline{v}_\alpha}}\right)} \label{eq:xi_s2_2} \\
        & = \left(\frac{\alpha}{\alpha+1}\right)^4  \nonumber \\
        & \quad \cdot \frac{\sum_{t=0}^{\alpha-1}\binom{l_{\alpha+1}}{t}\binom{\alpha+1}{t}\left\{\binom{l_{\alpha+1}}{\alpha+1}^{-\frac{t}{\alpha}}-\binom{l_{\alpha+1}}{\alpha+1}^{-1}\right\}}
        {\sum_{t=0}^{\alpha-2}\binom{l_{\alpha}}{t}\binom{\alpha}{t}\left\{\binom{l_{\alpha}}{\alpha}^{-\frac{t}{\alpha-1}} - \binom{l_{\alpha}}{\alpha}^{-1}\right\}} \label{eq:xi_s2_3} \\
        & \approx \left(\frac{\alpha}{\alpha+1}\right)^4 \cdot \frac{\sum_{t=0}^{\alpha-1}\binom{l_{\alpha+1}}{t}\binom{\alpha+1}{t}\binom{l_{\alpha+1}}{\alpha+1}^{- \frac{t}{\alpha}}}
        {\sum_{t=0}^{\alpha-2}\binom{l_{\alpha}}{t}\binom{\alpha}{t}\binom{l_{\alpha}}{\alpha}^{-\frac{t}{\alpha-1}}}, \label{eq:xi_s2_5}
   	\end{align}
    where \eqref{eq:xi_s2_2} and \eqref{eq:xi_s2_3} follow from Lemma~\ref{thm:v_ratio} and \eqref{eq:power_k_v}, respectively. Also, \eqref{eq:xi_s2_5} follows from 
    \begin{equation} \label{eq:xi_s2_6}
        \binom{l_{\alpha}}{\alpha}^{-\frac{t}{\alpha - 1}} \gg \binom{l_{\alpha}}{\alpha}^{-1}
    \end{equation}
    for $t \le \alpha - 2$ and a large $L$. We derive \eqref{eq:xi_s2_6} as follows:
    \begin{align}
        \binom{l_{\alpha}}{\alpha}^{-\frac{t}{\alpha - 1}} & \ge \binom{l_{\alpha}}{\alpha}^{-1} \cdot \binom{l_{\alpha}}{\alpha}^{\frac{1}{\alpha - 1}} \label{eq:xi_s2_7} \\
        & > \binom{l_{\alpha}}{\alpha}^{-1} \cdot \left( \frac{l_\alpha}{\alpha} \right)^{1 + \frac{1}{\alpha - 1}} \label{eq:xi_s2_8} \\
        & \gg \binom{l_{\alpha}}{\alpha}^{-1}, \label{eq:xi_s2_9}
    \end{align}
    where \eqref{eq:xi_s2_7} follows from $t \le \alpha - 2$ and \eqref{eq:xi_s2_8} follows from $\binom{l_\alpha}{\alpha}>(\frac{l_\alpha}{\alpha})^\alpha$ for $l_\alpha>\alpha$. Also, \eqref{eq:xi_s2_9} holds because $l_\alpha \gg \alpha$. 
   	

   	Finally, we show that \eqref{eq:xi_s2_5} converges to $\left( \frac{\alpha}{\alpha + 1} \right)^4$. For a large $L$, 
   	$l_\alpha = \left \lfloor \frac{L}{\overline{v}_\alpha} \right\rfloor \approx \frac{L}{\overline{v}_\alpha}$ and $\binom{l_\alpha}{t}\approx \frac{l_\alpha^t}{t!}$. Then, 
        \begin{align}
         &\sum_{t=0}^{\alpha-1}\binom{l_{\alpha+1}}{t}\binom{\alpha+1}{t}\binom{l_{\alpha+1}}{\alpha+1}^{-\frac{t}{\alpha}} \nonumber \\ 
         & \qquad \approx \sum_{t=0}^{\alpha-1} \binom{\alpha+1}{t} \cdot \frac{(l_{\alpha+1})^{t(1-\frac{\alpha+1}{\alpha})}}{t! \cdot \{(\alpha+1)!\}^{-\frac{t}{\alpha}}} \\
         & \qquad \approx\sum_{t=0}^{\alpha-1} \binom{\alpha+1}{t} \cdot \frac{ \left(\frac{L}{\overline{v}_{\alpha+1}}\right)^{-\frac{t}{\alpha}}}{t! \cdot \{(\alpha+1)!\}^{-\frac{t}{\alpha}}} \\
         & \qquad = \sum_{t=0}^{\alpha-1}\binom{\alpha+1}{t} \cdot \frac{\{(\alpha+1) ! \cdot \overline{v}_{\alpha+1}\}^{\frac{t}{\alpha}}}{t!} \cdot L^{-\frac{t}{\alpha}}.  \label{eq:xi_beta}
        \end{align}
        Also, 
        \begin{align}
            &\sum_{t=0}^{\alpha-2}\binom{l_{\alpha}}{t}\binom{\alpha}{t}\binom{l_{\alpha}}{\alpha}^{-\frac{t}{\alpha-1}} \nonumber \\
            & \qquad \approx \sum_{t=0}^{\alpha-2}\binom{\alpha}{t} \cdot \frac{(\alpha! \cdot \overline{v}_{\alpha})^{\frac{t}{\alpha-1}}}{t!} \cdot L^{-\frac{t}{\alpha-1}}. \label{eq:xi_alpha}
        \end{align} 
        
        For a large $L$, \eqref{eq:xi_beta} converges to one because the highest degree of $L^{-\frac{t}{\alpha}}$ is zero by $t=0$. Similarly, \eqref{eq:xi_alpha} converges to one. Hence, $\xi(\alpha) \approx \left(\frac{\alpha}{\alpha+1}\right)^4$ for a large $L$.

    
    

	
	\bibliographystyle{IEEEtran}
	\bibliography{abrv,mybib}

\begin{thebibliography}{10}
\providecommand{\url}[1]{#1}
\csname url@samestyle\endcsname
\providecommand{\newblock}{\relax}
\providecommand{\bibinfo}[2]{#2}
\providecommand{\BIBentrySTDinterwordspacing}{\spaceskip=0pt\relax}
\providecommand{\BIBentryALTinterwordstretchfactor}{4}
\providecommand{\BIBentryALTinterwordspacing}{\spaceskip=\fontdimen2\font plus
\BIBentryALTinterwordstretchfactor\fontdimen3\font minus
  \fontdimen4\font\relax}
\providecommand{\BIBforeignlanguage}[2]{{%
\expandafter\ifx\csname l@#1\endcsname\relax
\typeout{** WARNING: IEEEtran.bst: No hyphenation pattern has been}%
\typeout{** loaded for the language `#1'. Using the pattern for}%
\typeout{** the default language instead.}%
\else
\language=\csname l@#1\endcsname
\fi
#2}}
\providecommand{\BIBdecl}{\relax}
\BIBdecl

\bibitem{Turan2018recommendation}
M.~S. Turan, E.~Barker, J.~Kelsey, K.~A. McKay, M.~L. Baish, and M.~Boyle,
  \emph{{Recommendation for the Entropy Sources Used for Random Bit
  Generation}}, NIST Special Publication 800-90B Std., Jan. 2018.

\bibitem{Hagerty2012entropy}
P.~Hagerty and T.~Draper, ``{Entropy bounds and statistical tests},'' in
  \emph{Proc. NIST Random Bit Generation Workshop}, Dec. 2012, pp. 1--28.

\bibitem{Kelsey2015predictive}
J.~Kelsey, K.~A. McKay, and M.~S. Turan, ``{Predictive models for min-entropy
  estimation},'' in \emph{Proc. Int. Workshop Cryptograph. Hardw. Embedded
  Syst. (CHES)}, Berlin, Heidelberg, Sep. 2015, pp. 373--392.

\bibitem{Amaki2013worst}
T.~Amaki, M.~Hashimoto, Y.~Mitsuyama, and T.~Onoye, ``{A worst-case-aware
  design methodology for noise-tolerant oscillator-based true random number
  generator with stochastic behavior modeling},'' \emph{{IEEE} Trans. Inf.
  Forensics Security}, vol.~8, no.~8, pp. 1331--1342, Aug. 2013.

\bibitem{Killmann2011proposal}
W.~Killmann and W.~Schindler, \emph{A proposal for: Functionality classes for
  random number generators}, German Federal Office for Information Security
  (BSI) Std., Rev.~2, Sep. 2011.

\bibitem{Rukhin2010statistical}
A.~Rukhin, J.~Soto, J.~Nechvatal, M.~Smid, E.~Barker, S.~Leigh, M.~Levenson,
  M.~Vangel, D.~Banks, A.~Heckert, J.~Dray, and S.~Vo, \emph{{A statistical
  test suite for random and pseudorandom number generators for cryptographic
  applications}}, NIST Special Publication 800-22 Std., Rev.~1a, Apr. 2010.

\bibitem{Ben-Bassat1978renyi}
M.~Ben-Bassat and J.~Raviv, ``{Renyi's entropy and the probability of error},''
  \emph{{IEEE} Trans. Inf. Theory}, vol.~24, no.~3, pp. 324--331, May 1978.

\bibitem{Zhu2020analysis}
S.~Zhu, Y.~Ma, X.~Li, J.~Yang, J.~Lin, and J.~Jing, ``{On the analysis and
  improvement of min-entropy estimation on time-varying data},'' \emph{{IEEE}
  Trans. Inf. Forensics Security}, vol.~15, pp. 1696--1708, Oct. 2020.

\bibitem{Beck1993thermodynamics}
C.~Beck and F.~Sch{\"{o}}gl, \emph{{Thermodynamics of Chaotic Systems: An
  Introduction}}, ser. Cambridge Nonlinear Science Series.\hskip 1em plus 0.5em
  minus 0.4em\relax Cambridge University Press, 1993.

\bibitem{Goldreich2000testing}
O.~Goldreich and D.~Ron, ``{On testing expansion in bounded-degree graphs},''
  \emph{Electron. Colloq. Comput. Complexity}, vol.~7, Jan. 2000.

\bibitem{Batu2013testing}
T.~Batu, L.~Fortnow, R.~Rubinfeld, W.~D. Smith, and P.~White, ``{Testing
  closeness of discrete distributions},'' \emph{J. ACM}, vol.~60, no.~1, pp.
  4:1--4:25, Feb. 2013.

\bibitem{Acharya2015complexity}
J.~Acharya, A.~Orlitsky, A.~T. Suresh, and H.~Tyagi, ``{The complexity of
  estimating R{\'{e}}nyi entropy},'' in \emph{Proc. Annu. {ACM-SIAM} Symp.
  Discrete Algorithms (SODA)}, Jan. 2015, pp. 1855--1869.

\bibitem{Zhu2017analysis}
S.~Zhu, Y.~Ma, T.~Chen, J.~Lin, and J.~Jing, ``{Analysis and improvement of
  entropy estimators in NIST SP 800-90B for non-IID entropy sources},''
  \emph{IACR Trans. Symmetric Cryptol.}, vol. 2017, no.~3, pp. 151--168, Sep.
  2017.

\bibitem{Golic1987relationship}
J.~Golic, ``{On the relationship between the information measures and the Bayes
  probability of error},'' \emph{{IEEE} Trans. Inf. Theory}, vol.~33, no.~5,
  pp. 681--693, Sep. 1987.

\bibitem{Bar-Yossef2001sampling}
\BIBentryALTinterwordspacing
Z.~Bar-Yossef, R.~Kumar, and D.~Sivakumar, ``{Sampling algorithms: Lower bounds
  and applications},'' in \emph{Proc. Annu. {ACM} Symp. Theory Comput. (STOC)},
  Feb. 2002, pp. 266--275. [Online]. Available:
  \url{https://webee.technion.ac.il/people/zivby/papers/sampling/sampling_full.ps}
\BIBentrySTDinterwordspacing

\bibitem{Kim2021efficient}
Y.~Kim, C.~Guyot, and Y.-S. Kim, ``{On the efficient estimation of
  min-entropy},'' \emph{{IEEE} Trans. Inf. Forensics Security}, vol.~16, pp.
  3013--3025, Apr. 2021.

\bibitem{Reched2001renyi}
Z.~Rached, F.~Alajaji, and L.~{Lorne Campbell}, ``{Renyi's divergence and
  entropy rates for finite alphabet Markov sources},'' \emph{{IEEE} Trans. Inf.
  Theory}, vol.~47, no.~4, pp. 1553--1561, May 2001.

\bibitem{Kamath2016estimation}
S.~Kamath and S.~Verd{\'{u}}, ``{Estimation of entropy rate and R{\'{e}}nyi
  entropy rate for Markov chains},'' in \emph{Proc. IEEE Int. Symp. Inf. Theory
  (ISIT)}, Jul. 2016, pp. 685--689.

\bibitem{Random}
\BIBentryALTinterwordspacing
``{RANDOM.ORG}.'' [Online]. Available: \url{https://www.random.org}
\BIBentrySTDinterwordspacing

\bibitem{Ubld}
\BIBentryALTinterwordspacing
``{Ubld.it: TrueRNG}.'' [Online]. Available:
  \url{http://ubld.it/products/truerng-hardware-random-number-generator/}
\BIBentrySTDinterwordspacing

\end{thebibliography}

\end{document}